\documentclass[english]{elsarticle}
\usepackage[T1]{fontenc}
\usepackage[latin9]{inputenc}
\usepackage{color}
\usepackage{array}
\usepackage{float}
\usepackage{multirow}
\usepackage{amsmath}
\usepackage{graphicx}

\makeatletter

\providecommand{\tabularnewline}{\\}
\floatstyle{ruled}
\newfloat{algorithm}{tbp}{loa}
\providecommand{\algorithmname}{Algorithm}
\floatname{algorithm}{\protect\algorithmname}

\usepackage{fullpage}
\usepackage{url}

\makeatother

\usepackage{babel}
\usepackage{listings}

\begin{document}
\title{Evaluation of the Partitioned Global Address Space (PGAS) model for an inviscid Euler solver} 
\author[uibk]{Martina Prugger\corref{cor1}}
\ead{martina.prugger@uibk.ac.at}
\cortext[cor1]{Corresponding author}
\author[uibk]{Lukas Einkemmer} 
\author[uibk]{Alexander Ostermann}

\address[uibk]{Department of Mathematics, University of Innsbruck, Austria} 

\begin{abstract} In this paper we evaluate the performance of Unified Parallel C (which implements the partitioned global address space programming model) using a numerical method that is widely used in fluid dynamics. In order to evaluate the incremental approach to parallelization (which is possible with UPC) and its performance characteristics, we implement different levels of optimization of the UPC code and compare it with an MPI parallelization on four different clusters of the Austrian HPC infrastructure (LEO3, LEO3E, VSC2, VSC3) and on an Intel Xeon Phi. We find that UPC is significantly easier to develop in compared to MPI and that the performance achieved is comparable to MPI in most situations. The obtained results show worse performance (on VSC2), competitive performance (on LEO3, LEO3E and VSC3), and superior performance (on the Intel Xeon Phi) compared with MPI. \end{abstract} 
\maketitle

\section{Introduction}

Both in industry and in academia, fluid dynamics is an important research
area. Lots of scientific codes have been developed that predict weather
patterns, simulate the behavior of flows over aircrafts, or describe
the density within interstellar nebulae.

Usually, such codes compute the numerical solution of an underlying
partial differential equation (PDE) describing the phenomenon in question.
Describing the time evolution of a fluid typically leads to nonlinearities
of the governing PDEs. A classic example of such a system of equations
are the so-called Euler equations of gas dynamics. They are used to
describe compressible, inviscid fluids by modeling the behavior of
mass density, velocity, and energy. An important aspect of these equations
is that due to the nonlinearity so-called shock waves (i.e., rapid
changes in the medium) can form. Therefore, special numerical methods
have to be used that can cope with these discontinuities without diminishing
the performance. A number of software packages has been developed
that are used both in an industrial as well as in an academic setting.
However, there is still a lot of progress to be made with respect
to numerical integrators and their implementation on large scale HPC
systems.

A typical supercomputer consists of a number of connected computers
that work together as one system. To exploit such a system, parallel
programming techniques are necessary. The most important programming
model to communicate between the different processes within such a
cluster is message passing, which is usually implemented via the Message
Passing Interface (MPI) standard. 

In recent years, MPI has become the classical approach for HPC applications.
This is due to its portability between different hardware systems
as well as its scalability on large clusters. However, the standard
is mainly focused on two-sided communication, i.e., the transfer has
to be acknowledged by both the sender as well as the receiver of a
message. This is true even if both processes are located on the same
computation node and thus share the same memory. On today's HPC systems,
this overhead is not significant, however, experts predict that on
future generations of supercomputers, e.g. exascale systems, this
may result in a noticeable loss of performance. There are various
approaches to combine MPI code for off-node communication with OpenMP
for on-node communication into a hybrid model, however, the development
of such codes becomes more difficult.

Since parallelization with MPI has to be done explicitly by the programmer,
parallelizing a sequential code is in many situations quite difficult
(even without considering a hybridization with OpenMP). In recent
years Partitioned Global Address Space (PGAS) languages have emerged
as a viable alternative for cluster programming. PGAS languages like,
e.g., Unified Parallel C (UPC) or Coarray Fortran try to exploit the
principle of locality on a compute node. The syntax to access data
is similar to the OpenMP approach, which is usually easier for the
programmer than MPI (see, e.g., \cite{easier1}). However, in contrast
to OpenMP, it offers additional locality control (which is important
for distributed memory systems). Thus, a PGAS language is able to
naturally operate within a modern cluster (a distributed memory system
that is build from shared memory nodes).

The computer code can access data using one-sided communication primitives,
even if the actual data resides on a different node. The compiler
is responsible for optimizing data transfers (and thus, if implemented
well, can be as effective as MPI). However, usually a naive implementation
does not result in optimal performance. In this case the programmer
has the opportunity to optimize the parallel code step by step until
the desired level of scaling is achieved. For further information
about UPC, see, e.g., \cite{upc_info3,upc_info1,upc_info2}.

Since PGAS systems are an active area of research, there still may
occur problems with hardware compatibility as well as compiler optimization
and portability. Such issues usually do no longer affect MPI systems,
due to the time in development as well as the popularity of MPI.

Nevertheless, the PGAS approach seems to be a viable alternative for
researchers to develop parallel scientific codes that scale well even
on large HPC systems.

\subsection{Related work}

A significant amount of research has been conducted with respect to
the performance and the usability (for the latter see, e.g., \cite{prod})
of PGAS languages. Particularly the NAS benchmark is a popular collection
of test problems to explore the PGAS paradigm. For example, in \cite{benchmark1},
the NAS benchmark is used to investigate the UPC language, while \cite{benchmark2}
and \cite{bm_nas_cray} employ the NAS benchmark to compare UPC with
an MPI as well as an OpenMP parallelization (the latter on a vendor
supported Cray XT5 platforms). Another kind of benchmark to measure
fine- and course-grained shared memory accesses of UPC was developed
in \cite{bm_upc}. 

Even though benchmarks can give an interesting idea of how PGAS languages
behave for simple codes, additional problems can and do occur in more
involved programs. This leads to an investigation of the PGAS paradigm
in different fields. In \cite{involved1}, the mini-app CloverLeaf
\cite{cloverleaf}, which implements a Lagrangian-Euler scheme to
solve the two-dimensional Euler equations of gas dynamics is implemented
using two PGAS approaches. This paper extensively optimizes the implementation
using OpenSHMEM as well as Coarray Fortran. They manage to compete
with MPI for up to $50,000$ cores on their high end systems. In contrast,
our work considers HPC systems for which no vendor support for UPC
or Coarray Fortran is provided (while in the before mentioned paper,
a CRAY XC30 and an SGI ICE-X system are used). In addition, \cite{involved1}
is not concerned with the usability of UPC for medium sized parallelism
(which is a main focus of our work).

Another use of the PGAS paradigm in computational fluid dynamics (CFD)
is described in \cite{cfd_pgaslibrary}, where the unstructured CFD
solver TAU was implemented as a library using the PGAS-API GPI.

Let us also mention \cite{involved2}, where the old legacy high latency
code FEniCS (which implements a finite element approach on an unstructured
grid) is improved by using a hybrid MPI/PGAS programming model. Since
the algorithm used in the FEniCS code is formulated as a linear algebra
program, the above mentioned paper substitutes the linear algbra backend
PETSc with their own UPC based library. In contrast, in our work we
employ a direct implementation of a specific numerical algorithm (as
it is well known that a significant performance penalty is paid, if
this algorithm on a structured grid is implemented using a generic
linear algebra backend) and focus on UPC as a tool to simplify parallel
programming (while \cite{involved2} uses UPC in order to selectively
replace two-sided communication with one-sided communication to make
a legacy application ready for exascale computing).

The use of PGAS languages has also been investigated on the Intel
Xeon Phi. In \cite{xeonphi_openshmem}, e.g., the performance of OpenSHMEM
is explored on Xeon Phi clusters. Furthermore, \cite{xeonphi_upc}
implements various benchmarks including the NAS benchmark with UPC
on the Intel Xeon Phi. However, to the best of our knowledge, no realistic
application written in UPC has been investigated on the Intel Xeon
Phi.

\subsection{Research goals}

The goal of this paper is to investigate the performance of a fluid
solver that is based on a widely used numerical algorithm for solving
the Euler equations of gas dynamics. In contrast to \cite{involved1},
we use a pure Eulerian approach that uses a Riemann problem. More
accurate methods have been derived from this basic technique (see,
e.g., \cite{eulergodu1,eulergodu2}) and we thus consider this method
as a baseline algorithm for future optimizations. The details of this
method are described in the next section. 

This paper tries to give an idea of the usability of UPC for medium-sized
not vendor-supported systems. In that respect, we used four HPC systems
of the Austrian research infrastructure (VSC2, VSC3, LEO3, LEO3E,
the Berkeley UPC \cite{berkeley_compiler} package is used on each
of these systems). We believe that these systems are typical in terms
of the HPC resources that most researchers have available and that
many practitioners could profit from the programming model that UPC
(and PGAS languages in general) offer.

To demonstrate the usability of the PGAS paradigm, we implemented
our algorithm with UPC as well as MPI. We are especially interested
in how the performance of UPC improves as we provide a progressively
better optimized code. Thus, the parallelization is done step by step
and the scalability of these versions is investigated.

In the next section, we introduce the basics of the sequential program
and describe its parallelization. Then, we describe the results we
obtained on different systems.

\section{Implementation and Parallelization }

We consider the Euler equations of gas dynamics in two space dimensions,
i.e.
\[
U_{t}+F(U)_{x}+G(U)_{y}=0,
\]
where 

\[
U=\left(\begin{array}{c}
\rho\\
\rho u\\
\rho v\\
E
\end{array}\right)
\]
is the vector of the conserved quantities: density $\rho$, momentum
in the $x$-direction $\rho u$, momentum in the $y$-direction $\rho v$,
and energy $E$. The flux is given by

\[
F(U)=\left(\begin{array}{c}
\rho u\\
\rho u^{2}+p\\
\rho uv\\
u(E+p)
\end{array}\right),\qquad G(U)=\left(\begin{array}{c}
\rho v\\
\rho uv\\
\rho v^{2}+p\\
v(E+p)
\end{array}\right).
\]

The equation is in conserved form. It can also be expressed by the
physical variables: density $\rho$, velocity in the $x$-direction
$u$, velocity in the $y$-direction $v$, as well as pressure $p$,
due to the relation $E=\rho\cdot\left(\frac{u^{2}+v^{2}}{2}+\frac{p}{(\gamma-1)\rho}\right)$.
Here, $\gamma$ is a physical constant. Usually, numerical codes are
tested for an ideal gas, where $\gamma=1.4$. We also use that setup
in this paper.

These equations describe the time evolution of the conserved quantities
inside a closed system. Since there is no source or sink in the system,
the integrals of these variables are conserved. However, mass can
be transferred within the system in the $x$- and $y$-direction according
to the fluxes $F(u)$ and $G(u)$, respectively. We note that the
flux terms include nonlinearities that can lead to the development
of shock waves, even for smooth initial data. Shocks are important
physical phenomena and thus need to be captured accurately by the
numerical scheme. 

In this paper, we use the well known first order finite volume Godunov
method in one space dimension. To apply this scheme, we split our
problem into two one-dimensional problems. Godunov's method relies
on the fact that the Riemann Problem for a one-dimensional conservation
law 
\[
U_{t}+F(U)_{x}=0,
\]
with the initial values 
\[
U(x,0)=\begin{cases}
U_{L}, & x<0,\\
U_{R,} & x>0,
\end{cases}
\]
where $U_{L}$ and $U_{R}$ are constants, can be solved exactly. 

\begin{figure}
\begin{centering}
\includegraphics[scale=0.5]{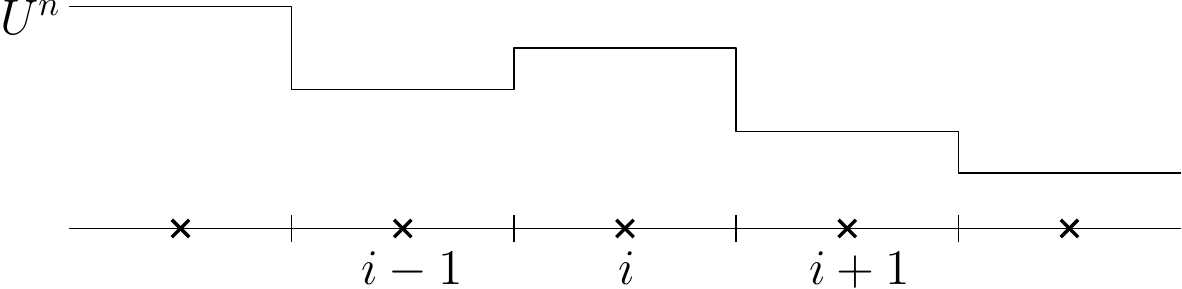}
\par\end{centering}

\protect\caption{Finite volume discretization at time step $n$. The conserved quantity
$U$ is averaged over each cell. At the cell interfaces, Riemann problems
occur. \label{fig:finite_volume_disc}}
\end{figure}

For the numerical method, we discretize our conserved quantities and
obtain $U_{i}^{n}$ at time step $n$ and grid point $i$ (see Figure
\ref{fig:finite_volume_disc}). 
 The value between the cell interfaces (located at $i-\frac{1}{2}$
and $i+\frac{1}{2}$) is constant. Thus at each cell interface a Riemann
problems occurs. We then solve these Riemann problems exactly in order
to update the value at the next time step. Godunov's method can then
be written as 
\[
U_{i}^{n+1}=U_{i}^{n}+\frac{\Delta t}{\Delta x}\left(F_{i-\frac{1}{2}}-F_{i+\frac{1}{2}}\right),
\]
 where $\Delta t$ and $\Delta x$ are the time step and the length
of a cell respectively. The numerical fluxes $F_{i-\frac{1}{2}}$
and $F_{i+\frac{1}{2}}$ are obtained by the Riemann solver at the
cell interface to the left and to the right of the grid point $i$.
Note that this method is restricted by a CFL condition. Since we are
mostly interested in the method and its parallelization, we choose
a fixed time step that is small enough such that the CFL condition
in our simulations is always satisfied. 

Since an exact solution can only be calculated for one spatial dimension,
we use Lie splitting to separate our two-dimensional problem into
two one-dimensional problems, i.e. we alternatingly solve:

\[
x\mbox{-direction: }\begin{cases}
U_{t}+F(U)_{x}=0\\
U(0)=U^{n}
\end{cases}\overset{\Delta t}{\Rightarrow}U^{n+\frac{1}{2}}=U(\Delta t)
\]

and 
\[
y\mbox{-direction: }\begin{cases}
U_{t}+G(U)_{y}=0\\
U(0)=U^{n+\frac{1}{2}}
\end{cases}\overset{\Delta t}{\Rightarrow}U^{n+1}=U(\Delta t).
\]

\medskip{}

We first take a step into the $x$-direction (using Godunov's method)
and then take a step into the $y$-direction to conclude our time
step. 

For a more detailed discussion of the Euler equations and Godunov's
method, see, e.g., \cite{eulergodu2,eulergodu1,toro2009riemann}.

\begin{figure}[h]
\begin{centering}
\includegraphics[width=14cm]{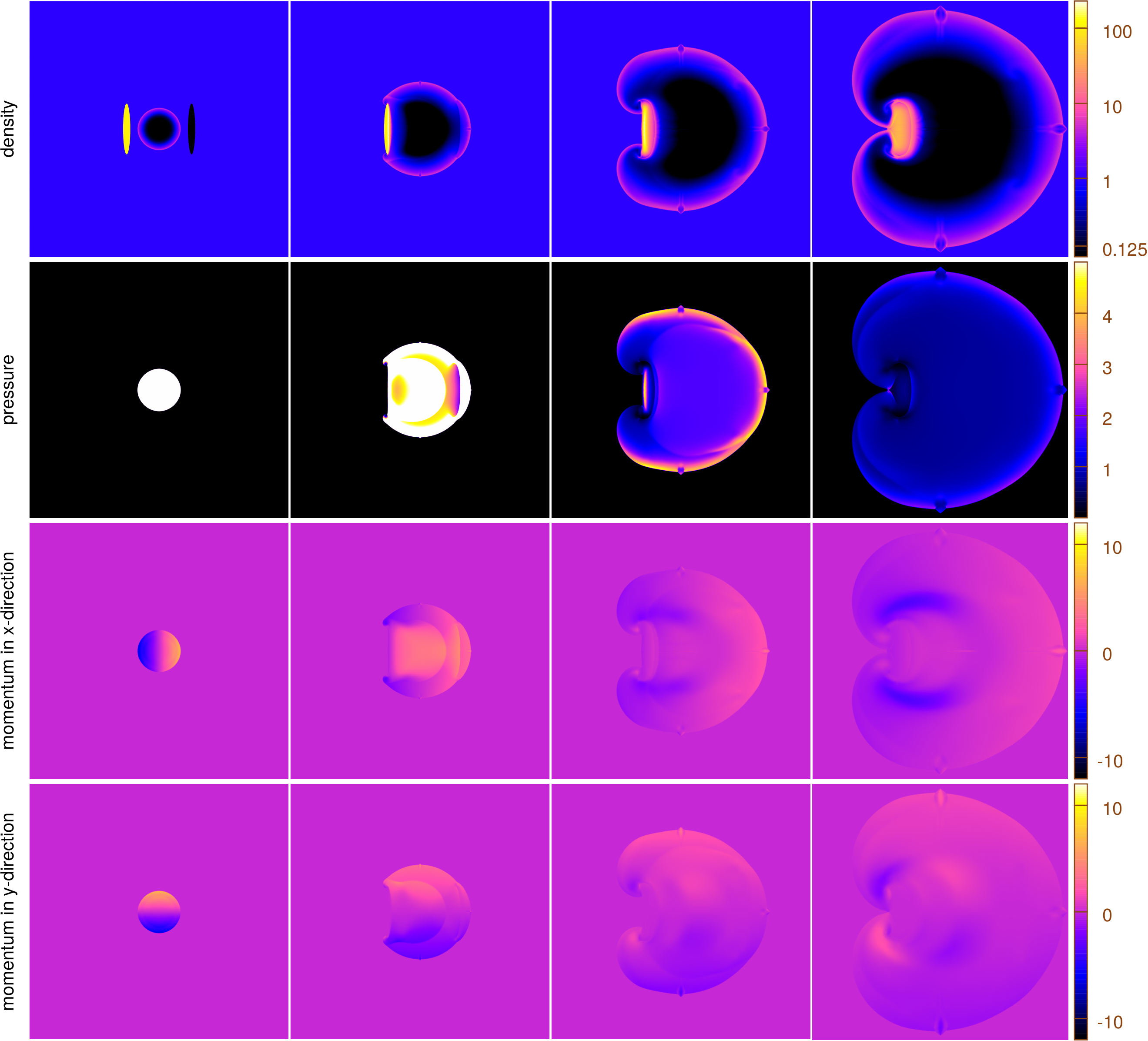}
\par\end{centering}

\protect\caption{Simulation of a Sedov explosion on the domain $[0,1]\times[0,1]$.
The variables are shown from top to bottom in the following order:
density, pressure, velocity in the $x$- and velocity in the $y$-direction.
From left to the right, snapshots in time are shown at $t=0.005$,
$t=0.025$, $t=0.075$ and $t=0.15$. The initial condition is chosen
as follows: the velocity is zero in both directions, the density is
set to one except for a hill of 100 at the left and a basin of 0.01
at the right. The pressure is a sharp Gaussian distribution. The density
forms a shock front which wraps around the density hill and is accelerated
by the sink.\label{fig:Simulation}}
\end{figure}

Figure \ref{fig:Simulation} shows the behavior of the physical variables
for a simple test problem on the domain $[0,1]\times[0,1]$. The snapshots
are taken at time $t=0.005$, $t=0.025$, $t=0.075$, and $t=0.15$.
As initial condition, we set a background density to $\rho=1$ with
a density hill of $\rho=100$ on the left side and a density basin
of $\rho=0.01$ on the right side of the domain. The pressure has
a background value of $p=0.0001$ and a Gaussian with a peak of $1200$
and the standard deviation $\sigma=20.48$. Both the velocity in the
$x$- and in the $y$-direction are $0$ at the beginning of the simulation.

Up to the time considered in the first column of Figure \ref{fig:Simulation},
the dynamics of the system is mainly determined by the variation in
pressure (as the initial values for the other variables are constant).
For the density field, a sharp front develops that spreads out symmetrically.
This is the propagation of a shock front which we have discussed earlier.
Note that the density variable varies over three orders of magnitude.
Thus, we use a logarithmic scale in order to better observe the behavior
of the solution. Such a scaling is not necessary for the other variables.
The pressure flattens out very quickly during this time period where
it causes an acceleration of the particles. To be able to observe
the behavior of the pressure variable at the last shown time step,
we limited the color range to a maximum of $5$ (the largest value
is close to $49$; at time $t=0.005$).

In the second column, we see that the density shock wave hits the
hill and the basin of the density field. For the hill, we observe
that the shock wave is absorbed by the much higher density, while
at the basin, it is drawn into the region of lower pressure. This
behavior can be observed by the pressure variable as well. When we
integrate further, the original shock wave wraps around the hill,
until it completely envelopes the hill and vortices begin to form.
The velocity variables show that the particles within the right part
of the domain are further accelerated, while at the hill, there is
almost no movement.

\subsection{Structure of the sequential code}

We start with a sequential code which we will parallelize both with
MPI and UPC. The basic structure of the code is described in Algorithm
\ref{alg:sequ}.

\begin{algorithm}[h]
\begin{lstlisting}
set initial physical variables;
time loop
  call_one_d in x-direction;
  call_one_d in y-direction;
\end{lstlisting}
 with the function 
\begin{lstlisting}
call_one_d:
  change physical to conserved variables;
  calculate the flux;
  execute Godunov's method;
  change conserved to physical variables;
\end{lstlisting}

\protect\caption{Pseudo code of the sequential implementation.\label{alg:sequ}}
\end{algorithm}

\subsection{Parallelization: MPI}

The MPI version of the code is based on the Single Program Multiple
Data (SPMD) approach. The two-dimensional arrays for the four physical
variables $\rho$, $u$, $v$ and $p$ are partitioned among the number
of processes in a row-wise manner. In this simulation, we take a look
at two different boundary conditions. In the $x$-direction, we use
periodic boundary conditions, i.e., the left most grid point is identified
with the right most grid point. In the $y$-direction, we consider
an inflow condition at the bottom (i.e., the value at the bottom is
a constant that we define as a global variable) and an outflow condition
at the top (we just use the value on the top of the grid twice, which
corresponds to Neumann boundary conditions). Since such a setting
will lead to unintended reflections, we introduce a couple of ghost
cells to hide these artifacts.

\begin{figure}[h]
\begin{centering}
\includegraphics[scale=0.5]{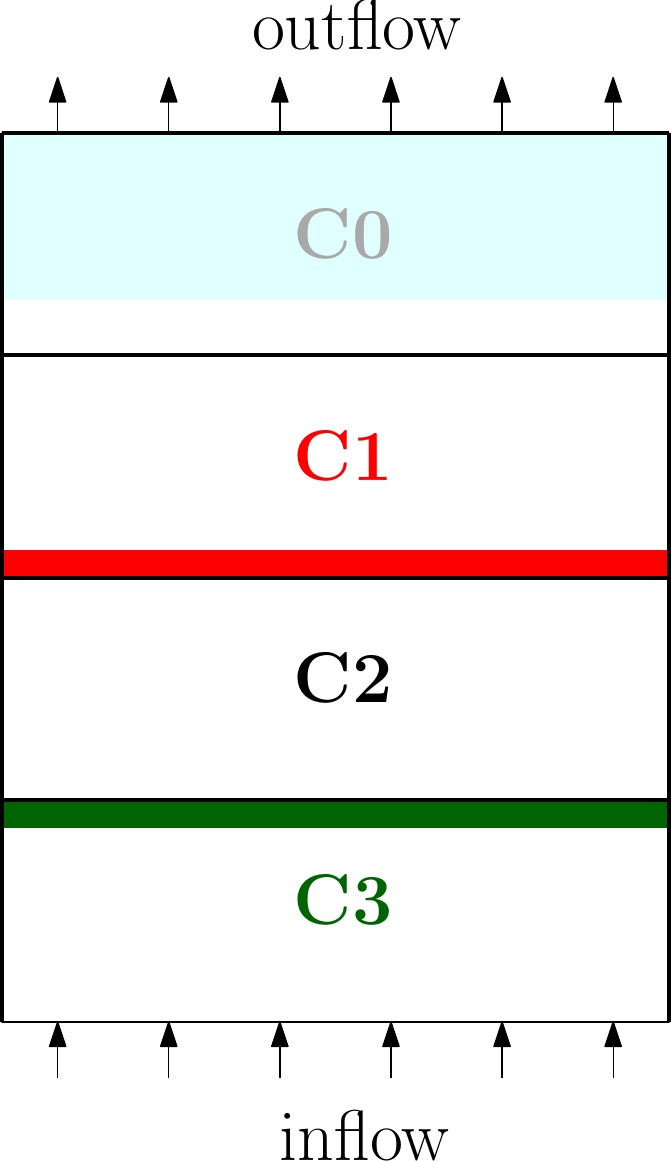}
\par\end{centering}

\protect\caption{Setup for the core C2, where the physical variables are stored row-wise
on the different cores. In the MPI implementation, the neighbors C1
in red and C3 in green prepare their bottom and top row respectively
and send it via \textit{MPI\_Isend} to core C2. In the \textit{naive},\textit{
pointer}, and\textit{ barrier} UPC implementations, C2 accesses these
rows directly via a shared array. In the UPC \textit{halo} implementation,
C2 fetches these rows via \textit{upc\_memget} without active involvement
from the neighbors. Due to the outflow condition, additional ghost
cells are necessary which are shown in cyan. Note that these cells
can extend over more than a single thread. The setup is valid for
the MPI \textit{row} implementation as well as the \textit{naive},
\textit{pointer}, \textit{barrier}, and \textit{halo} UPC implementations.
\label{fig:row-setup}}
\end{figure}

\begin{figure}[h]
\begin{centering}
\includegraphics[scale=0.63]{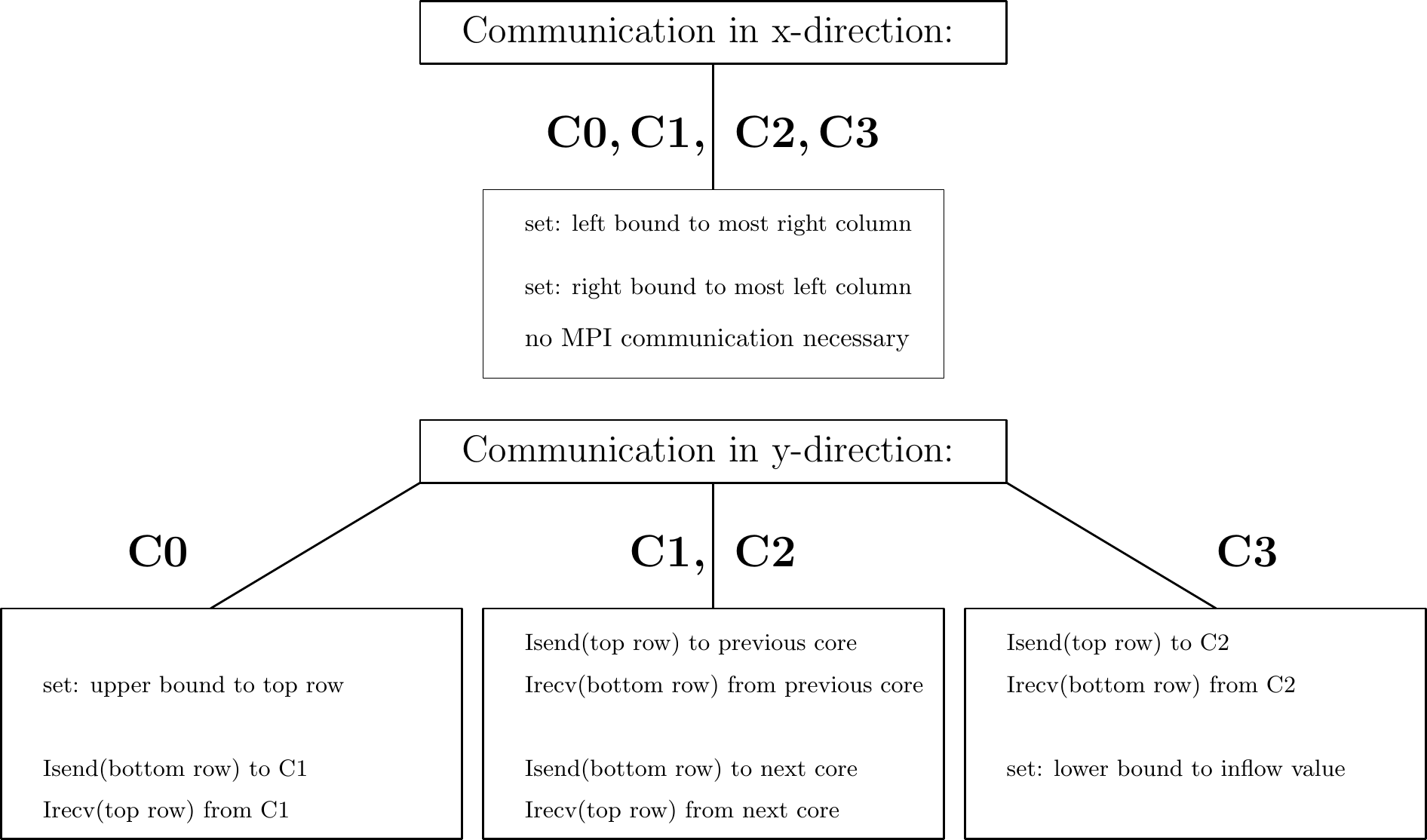}
\par\end{centering}

\protect\caption{Required communication diagram for the MPI \textit{row} setup on four
cores. \label{fig:MPI-row-diagram}}

\end{figure}

First, let us consider the partitioning of the domain in rows. This
situation is illustrated in Figure \ref{fig:row-setup}. Due to the
imposed boundary conditions, no MPI-communication is needed at the
boundary of the domain. However, each process has to communicate with
the two processes which are located immediately above and below of
it (this is illustrated in Figure \ref{fig:row-setup} and in more
detail in Figure \ref{fig:MPI-row-diagram} for a simulation with
four cores). Each process uses the non-blocking \textit{MPI\_Isend}
and \textit{MPI\_Irecv} routines in order to transfer the necessary
data. Note that no global synchronization is necessary.

For comparison, we also consider the distribution of the data in patches.
This framework exchanges less data but requires more MPI calls. 

Since the data is no longer distributed row-wise on each core, both
in the $x$- as well as in the $y$-direction, MPI-communication has
to be performed using the non-blocking MPI routines. Similar to the
row-wise decomposition, only communication with nearest neighbors
is required.

\begin{figure}[h]
\begin{centering}
\includegraphics[scale=0.5]{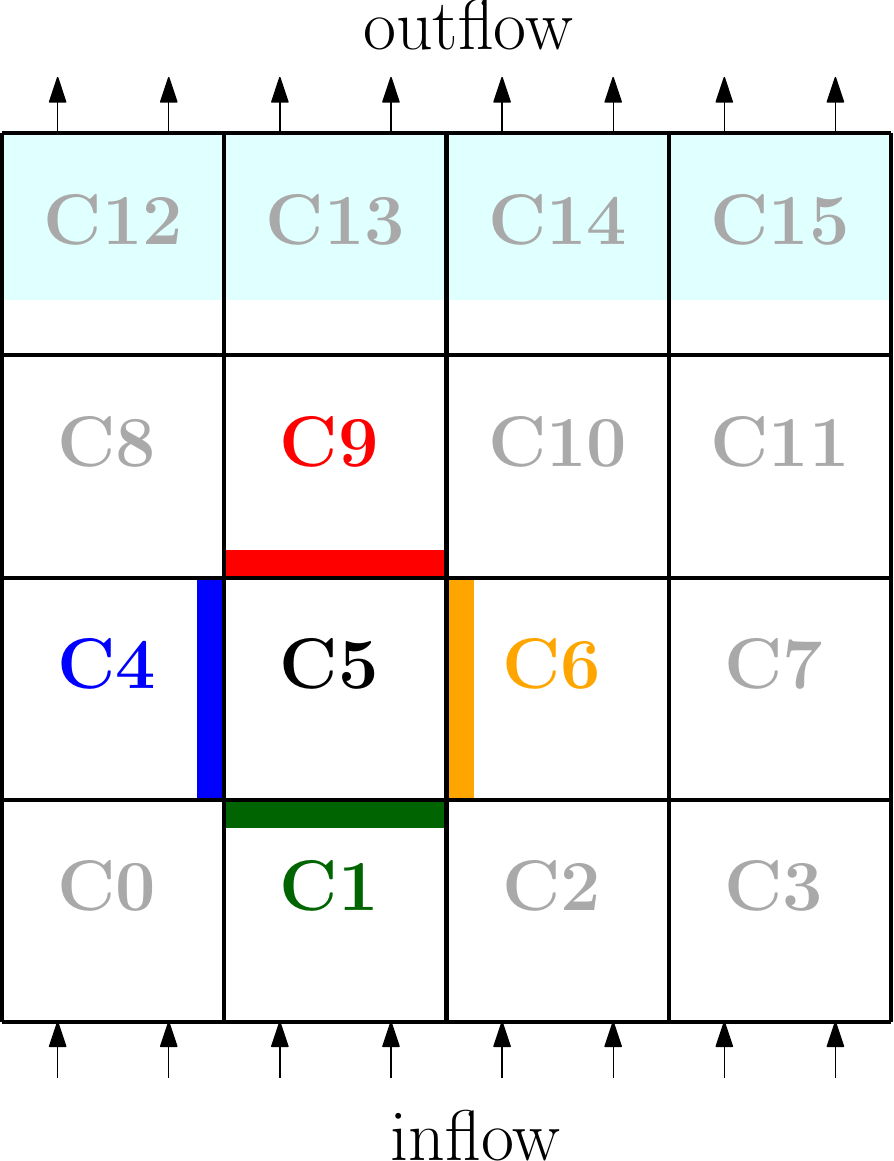}
\par\end{centering}

\protect\caption{Setup for the core C5, where the physical variables are stored patch-wise
on the different cores. In the MPI implementation, the neighbors C6
in orange, C9 in red, C4 in blue and C1 in green prepare their left,
bottom, right and top row respectively and send it via \textit{MPI\_Isend}
to core C5. In the UPC implementation, C5 fetches these rows via \textit{upc\_memget}
without active involvement of the neighbors. Due to the outflow condition,
additional ghost cells are necessary which are shown in cyan. Note
that these cells can extend over more than a single thread. The setup
is valid for the MPI \textit{patch} implementation as well as the
\textit{patch} UPC implementation.\label{fig:patch-setup}}
\end{figure}

\begin{figure}[h]
\begin{centering}
\includegraphics[scale=0.63]{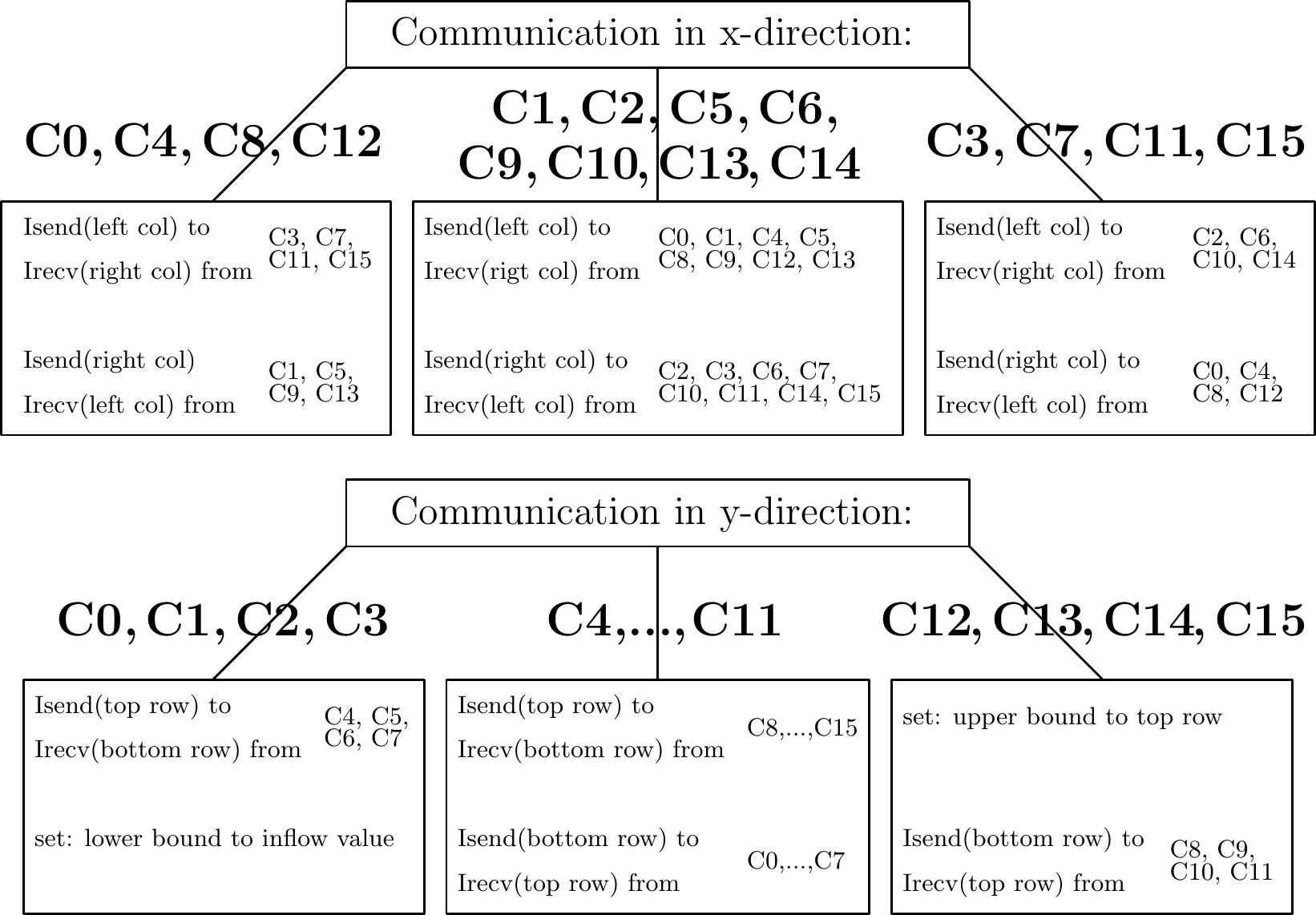}
\par\end{centering}

\protect\caption{Required communication diagram for the MPI \textit{patch} setup on
sixteen cores. \label{fig:patch-mpi-diagram}}

\end{figure}

In \ref{fig:patch-setup} we illustrate the partitioning of the data
on sixteen cores and highlight the required communication for the
core C5. In Figure \ref{fig:patch-mpi-diagram}, the communication
pattern of this setup is demonstrated in a diagram for sixteen cores.
For both implementations, the sequential code is extended with an
additional function that communicates the boundary values (see Algorithm
\ref{alg:MPI}).

\begin{algorithm}[h]
\begin{lstlisting}
set initial physical variables;
time loop
  communicate boundaries in x-direction;
  call_one_d in x-direction;
  communicate boundaries in y-direction;
  call_one_d in y-direction;
\end{lstlisting}

\protect\caption{Pseudo code of the MPI implementation.\label{alg:MPI}}
\end{algorithm}
In each time step, first the boundaries in the $x$-direction are
communicated and then the calculations in this direction are performed
locally on each processor. To conclude the time step, the same is
done in the $y$-direction.

\subsection{Parallelization: UPC}

For the UPC parallelization, we take advantage of the shared data
structures provided by UPC. As discussed earlier, a big advantage
of UPC is the incremental approach to parallelization. It is therefore
possible to parallelize the sequential code by just adding a few additional
lines to the sequential code. However, to obtain good scalability
we require a more sophisticated implementation. We therefore introduce
a sequence of optimization steps, whose performance we will analyze
in the next section on different HPC systems. 

For our first parallelization, which we call the \textit{naive} approach,
we just declare our work arrays as shared such that every processor
can access it. Therefore, no visible communication is performed. However,
the workload on each thread still has to be defined by the programmer.
The cells with affinity to certain threads are distributed in rows
similar to the MPI implementation that we discussed earlier (see Figure
\ref{fig:row-setup}). Due to the assumed memory layout of UPC, this
simplifies the implementation. Note that this data distribution is
also valid for the next three implementations. 

Since in the $x$-direction the slices are located on the same thread,
all calculations can be done locally. However, in the $y$-direction,
the cells are distributed among the threads. In this case, the communication
is performed directly on the shared array which hides the communication
(which is required to satisfy the remote access) from the user.

\begin{figure}[h]
\begin{centering}
\includegraphics[scale=0.63]{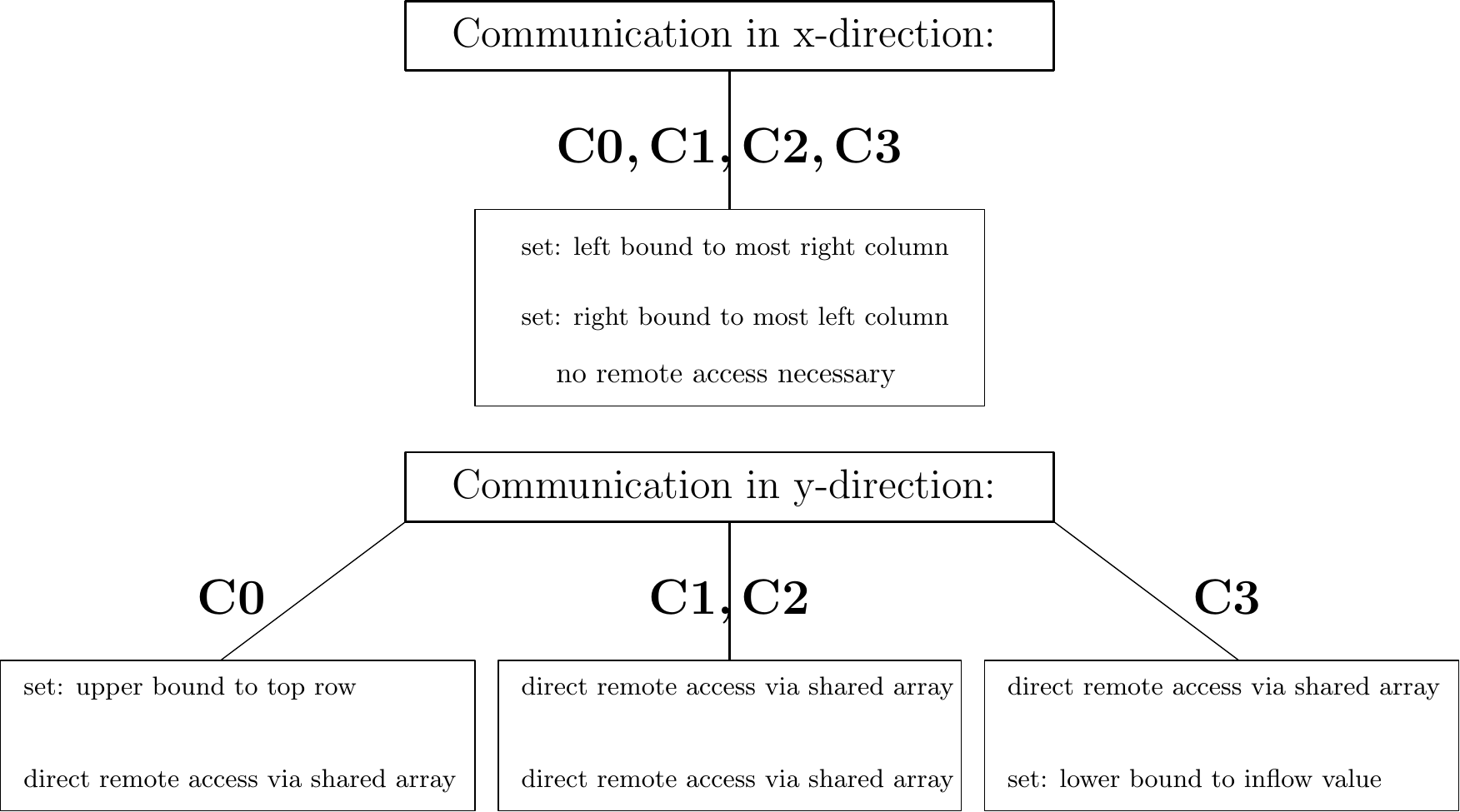}
\par\end{centering}

\protect\caption{Required communication diagram for the UPC: \textit{row}, \textit{pointer}
and \textit{barrier} setups on four cores. \label{fig:upc-naive-diagram}}

\end{figure}

Our code has the form stated in Algorithm \ref{alg:naive}. In Figure
\ref{fig:upc-naive-diagram}, the communication of this setup is demonstrated
in a diagram for four cores.

\begin{algorithm}[h]
\begin{lstlisting}
set shared initial physical variables;
time loop
  call_one_d in x-direction:
    do calculation;
  call_one_d in y-direction:
    use shared array for calculation;
\end{lstlisting}

\protect\caption{Pseudo code of the \textit{naive} UPC implementation.\label{alg:naive}}
\end{algorithm}

In this implementation, communication is the bottleneck. Even though
Thread 0 can access a data point on Thread 3, this might be very expensive,
especially, if both processes do not reside on the same node. As a
rule of thumb, shared variables are expensive to access, and thus
accessing them one by one should be avoided, if possible. In our application,
most of the memory access is local. 

We call our second approach \textit{pointer} approach, because in
the $y$-direction, we create a pointer on each thread that only manipulates
local data of the shared array slice, i.e. Algorithm \ref{alg:pointer}. 

\begin{algorithm}[h]
\begin{lstlisting}
set initial physical variables shared;
time loop
  call_one_d in x-direction:
    do calculation;
  call_one_d in y-direction:
    create local pointers to shared;
    use local pointers for calc;
\end{lstlisting}

\protect\caption{Pseudo code of the \textit{pointer} UPC implementation.\label{alg:pointer}}
\end{algorithm}

In this implementation, the communication is still performed indirectly,
by using the shared data array. The communication for sixteen cores
is demonstrated in Figure \ref{fig:upc-naive-diagram} and the data
distribution is still the same as above (see, e.g., Figure \ref{fig:row-setup}).
In principle, this optimization can be performed by the UPC compiler.
However, it is not clear, how efficient this optimization is in practice. 

Similar to the OpenMP programming paradigm, we need to avoid race
conditions, i.e., we need to make sure that when a calculation step
takes information from a shared object the corresponding data point
has already been updated. This is guaranteed by barriers. However,
at barriers all threads have to wait for each other until they can
continue with their work. This is accompanied with a significant overhead
and thus, barriers should be used as little as possible in the code. 

Using the performance tool \textit{upc\_trace} on our code we found
that the barriers in our code are a significant bottleneck. Our next
optimization step is therefore called the \textit{barrier} approach,
because we divided the calculation loop into two loops, so that we
can move the barrier outside of the loop. The idea is demonstrated
in Algorithm \ref{alg:barrier}. 

\begin{algorithm}[h]
\begin{lstlisting}
call_one_d in y direction:

define local calculation arrays glob;

start calculation loop
  cast local pointers;
  change physical to conserved variables in global array;
  calculate the flux in global array;
end calculation loop;
barrier;
start calculation loop
  cast local pointers;
  execute Godunov's method with global array;
  change conserved to physical variables;
end calculation loop;
\end{lstlisting}

\protect\caption{Pseudo code of the \textit{barrier} UPC implementation.\label{alg:barrier}}
\end{algorithm}

In this case, we have to define some arrays that we need for the calculation
globally (so that we can use them in the two loops). However, we only
have to call the barrier once in each time step, removing a lot of
overhead. Note that this approach only incurs a small overhead in
the amount of memory used. The distribution and communication in this
setup is the same as in the above method (see, e.g., Figure \ref{fig:row-setup}
and Figure \ref{fig:upc-naive-diagram}). 

Up to now, we only tried to exploit locality of the computation, but
kept the main working array shared. As the next step, we used an idea
which is usually implemented in MPI code: we do no longer work on
the whole shared array, but use shared variables only for communicating
ghost cells. Therefore, similar to MPI, the boundaries are communicated
in a single call to \textit{upc\_memget} and the remainder of the
computation is done locally. We thus name this level of optimization
the \textit{halo} approach (see Algorithm \ref{alg:halo}).

\begin{algorithm}[h]
\begin{lstlisting}
set initial physical variables shared;
get local copies of shared variables;
time loop
  call_one_d in x-direction:
    do calculation locally;
  barrier;
  get boundaries from shared array;
  barrier;
  call_one_d in y-direction:
    do calculation locally;
\end{lstlisting}

\protect\caption{Pseudo code of the \textit{halo} UPC implementation.\label{alg:halo}}
\end{algorithm}

\begin{figure}[h]
\begin{centering}
\includegraphics[scale=0.63]{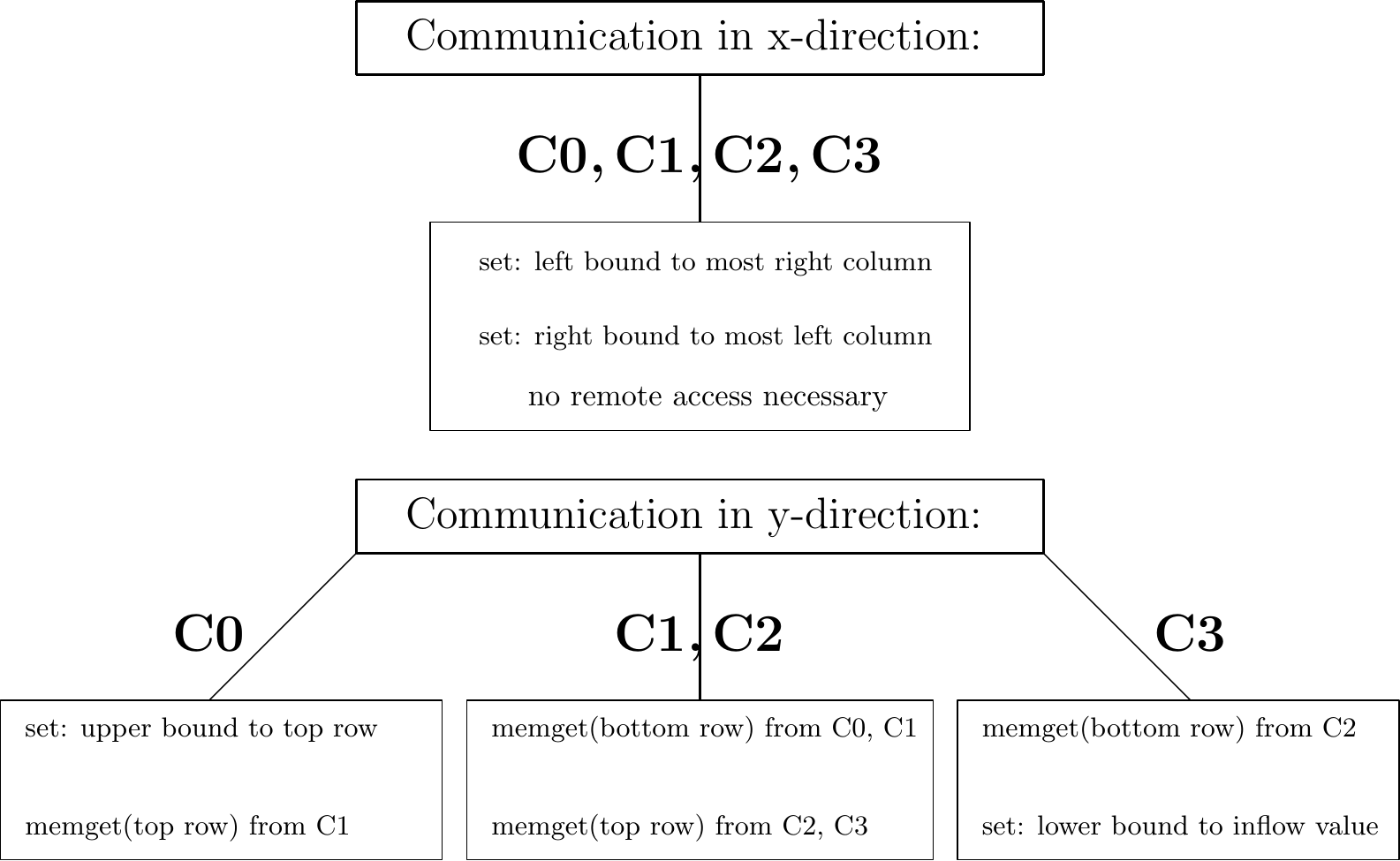}
\par\end{centering}

\protect\caption{Required communication diagram for the UPC: \textit{halo} setup on
four cores.\label{fig:upc-halo-diagram} }

\end{figure}

In this implementation, the computation is completely local. The call
\textit{upc\_memget} takes the necessary rows from the upper and lower
neighbor without them actively participating in the communication.
To avoid race conditions in this case, a synchronization barrier has
to be included. The distribution of the data is still equivalent to
the implementations above (see, e.g., Figure \ref{fig:row-setup}).
In Figure \ref{fig:upc-halo-diagram}, the communication of this setup
is demonstrated for four cores.

\begin{figure}
\begin{centering}
\includegraphics[scale=0.63]{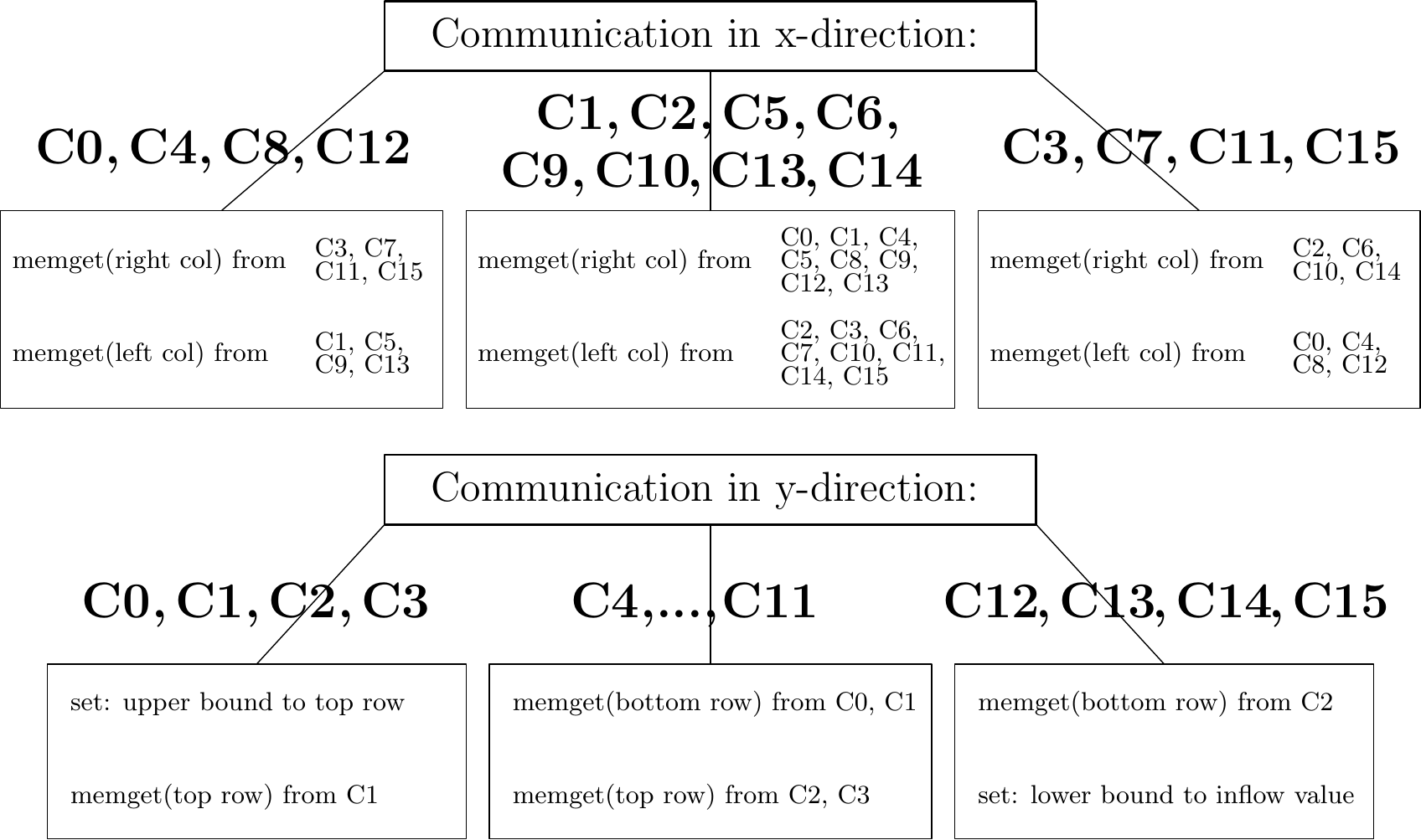}
\par\end{centering}

\protect\caption{Required communication diagram for the UPC: \textit{patch} setup on
sixteen cores.\label{fig:upc-patch-diagram}}

\end{figure}

All the approaches above are distributed in a row-wise manner. However,
this setup may lead to more communication than is necessary. A final
optimization step is thus to distribute the data points on the threads
in patches (see Figure \ref{fig:patch-setup}). Since the data points
are contiguous in the $x$-direction the data communication is done
by a single call of \textit{upc\_memget} (as was done in the previous
optimization step). To communicate the data in $y$-direction, we
used the strided function \textit{upc\_memget\_strided}. Note that
this function is part of the Berkeley UPC compiler and not yet part
of the UPC standard, however, it greatly facilitates the implementation.
Again, to avoid race conditions, synchronization barriers have to
be included. Due to the data distribution, we call this implementation
the \textit{patch} approach. Similar to the previous algorithm, this
setup performs only local computation. The communication for sixteen
cores is demonstrated in Figure \ref{fig:upc-patch-diagram}.

An overview for the different types of distribution, communication
and computation is given in Table \ref{tab:Overview}.

\begin{table}[h]
\begin{centering}
\begin{tabular}{clllll}
\hline 
 &  & distribution & computation & \multicolumn{1}{l}{comm: $x$-direction} & comm: $y$-direction\tabularnewline
 &  &  &  & left \& right boundary & top \& bottom boundary\tabularnewline
\hline 
\multirow{2}{*}{MPI} & \textit{row} & row-wise & local & none & Isend/Irecv\tabularnewline
 & \textit{patch} & patch-wise & local & Isend/Irecv & Isend/Irecv\tabularnewline
\hline 
\multirow{5}{*}{UPC} & \textit{naive} & row-wise & on shared array & none & implicit\tabularnewline
 & \textit{pointer} & row-wise & local, on shared array & none & implicit\tabularnewline
 & \textit{barrier} & row-wise & local, on shared array & none & implicit\tabularnewline
 & \textit{halo} & row-wise & local & none & upc\_memget\tabularnewline
 & \textit{patch} & patch-wise & local & upc\_memget & upc\_memget\tabularnewline
\hline 
\end{tabular}
\par\end{centering}

\protect\caption{Overview between different MPI and UPC implementations.\label{tab:Overview}}

\end{table}

As we can see, the optimization steps are getting more and more sophisticated
and the code gets more and more involved. However, changing and debugging
the code incrementally is usually much easier than writing an already
perfectly optimized version in the first place. The scaling behavior
of the different approaches on different hardware is discussed in
the next section.

\section{Results}

In this section we investigate the scaling behavior of the different
codes described in the previous section and compare the results on
different hardware. For this purpose, we have access to four different
HPC systems. LEO3 and LEO3E are local clusters at the University of
Innsbruck. Assembled in 2011 with 162 compute nodes, LEO3 is a medium
sized but relatively old cluster. In 2015, the computing infrastructure
in Innsbruck was extended by a smaller but more modern system, LEO3E,
with 45 computing nodes. Both systems have approximately equal peak
performance. 

In addition, we also use the main high performance facility in Austria:
the Vienna Scientific Cluster (VSC). In this study, we use VSC2, which
ranked 56 of the Top 500 systems when it came into operation in 2011
and consists of 1314 computing nodes. In addition, we consider VSC3,
which occupied rank 85 of the Top 500 systems in its first year of
operation (2014) and consists of 2020 computing nodes. 

We therefore have the opportunity to compare Tier 2 with Tier 1 systems
as well as relatively old with relatively new hardware.

All of the above systems are relatively traditional high performance
cluster models. We believe that such systems are representative for
the HPC resources available to most researchers. 

Since the shared memory model used by UPC is a natural fit for the
Intel Xeon Phi, we also investigate the scaling behavior of our implementation
on that platform. The Xeon Phi implements a many-core architecture
(similar to graphic processing units) with 60 physical cores (240
cores are available for hyperthreading). Since the cores of the Xeon
Phi are based on an x86 architecture, it is relatively straightforward
to compile a UPC program for it.

For detailed hardware specifications, see Table \ref{tab:Hardware}.

\begin{table}
\begin{centering}
\small%
\begin{tabular}{lrlrrrl}
system & nodes & CPU on node & cores & memory & Rpeak & nw controller\tabularnewline
\hline 
LEO3 & 162 & 2 x Intel Xeon X5650, 2.7 GHz & 12 & 4 TB & 18 TFlop/s & Mellanox\tabularnewline
LEO3E & 45 & 2 x Intel Xeon E5-2650-v3, 2.6 GHz & 10 & 4 TB & 29 TFlop/s & Mellanox\tabularnewline
VSC2 & 1314 & 2 x AMD Opteron 6132HE, 2.2GHz & 8 & 42 TB & 185 TFlop/s & Mellanox\tabularnewline
VSC3 & 2020 & 2 x Intel Xeon E5-2650v2, 2.6GHz & 8 & 131 TB & 682 TFlop/s & Intel\tabularnewline
\end{tabular}
\par\end{centering}

\protect\caption{Hardware specifications for the four clusters used in this paper.\label{tab:Hardware}}
\end{table}

On each system, we execute both a strong as well as a weak scaling
analysis. For the strong scaling analysis, we choose a fixed problem
size and run the program on different number of cores. Ideally, the
run time would decrease linearly in the number of cores. In this paper,
we choose our problem size as a grid of $512\times1024$ points and
a final time $T=0.005$. 

The weak scaling analysis is performed by increasing the grid according
to the number of cores. Ideally, the time it takes to finish the larger
problem would remain constant. In our case, we choose for one core
a domain with $64\times128$ grid points and a final integration time
of $T=0.001$. When quadrupling the number of cores, we quadruple
the problem size by choosing a grid of $128\times256$ for four cores,
etc. 

In our simulations we use a constant time step size $\Delta t=10^{-5}$
(which is small enough such that the CFL condition is always satisfied).

\subsection{Results on LEO3}

\begin{figure}[h]
\begin{centering}
\includegraphics[scale=0.72]{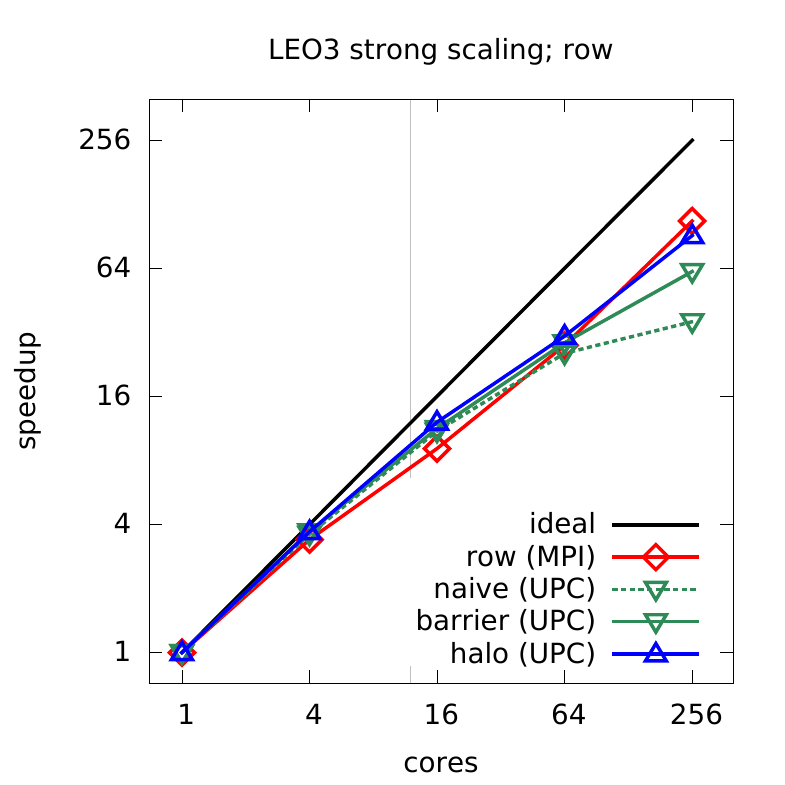}\includegraphics[scale=0.72]{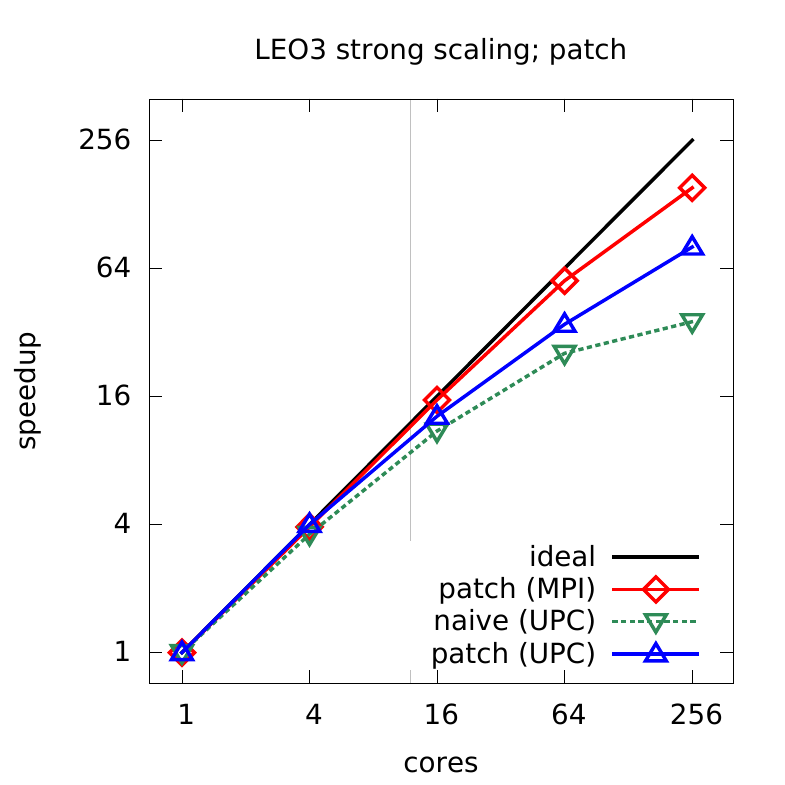}
\par\end{centering}

\begin{centering}
\includegraphics[scale=0.72]{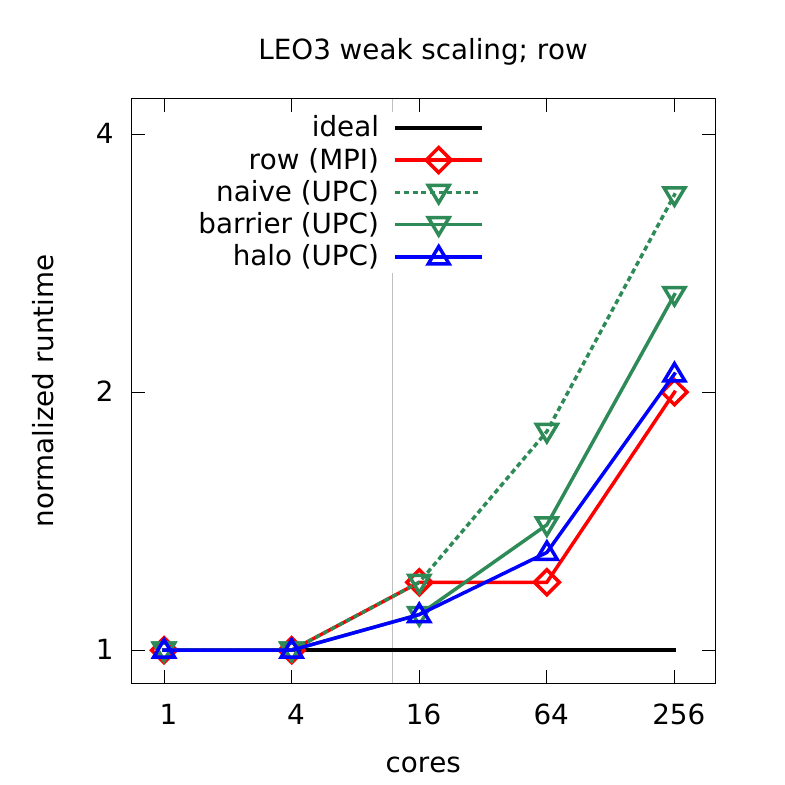}\includegraphics[scale=0.72]{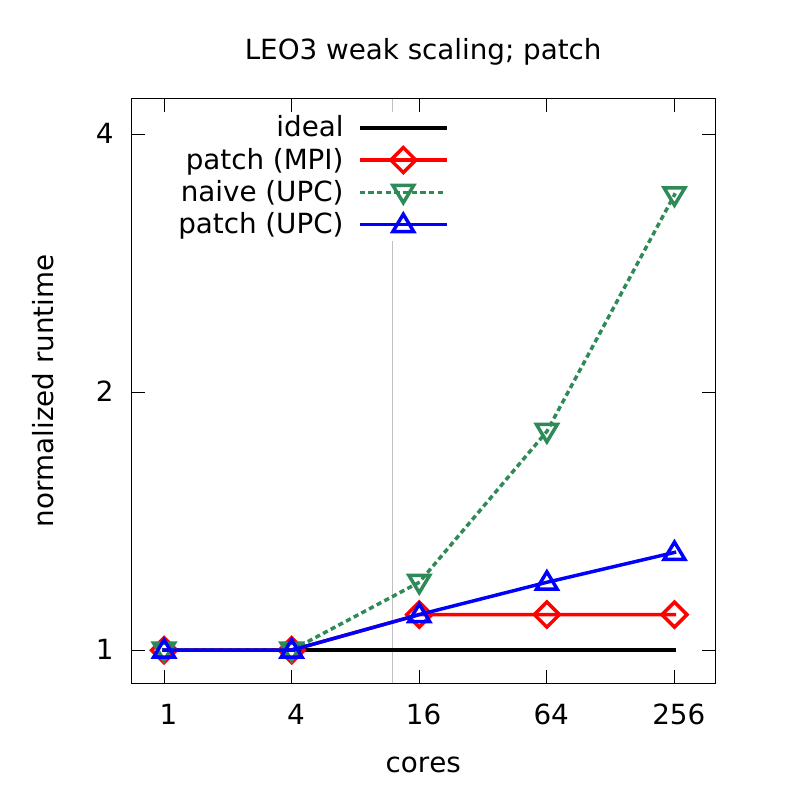}
\par\end{centering}

\protect\caption{This figure shows the scaling results on the LEO3. The left hand side
shows from top to bottom the strong and the weak scaling results for
the \textit{row-wise} communication pattern. Due to the similar results
of the UPC \textit{row} and \textit{pointer} version, we only include
the first one in the plot. The right hand side shows from top to bottom
the strong and the weak scaling results for the \textit{patched} communication
pattern. For comparison, this column also includes the \textit{row-wise}
distributed \textit{naive} UPC version. The black line represents
the ideal scaling in all cases. The gray line represents the number
of cores where the simulation can be run on a single node. For the
LEO3, this line is at 12 cores. \label{fig:Leo3}}
\end{figure}

On LEO3 we use up to 256 cores. Even though the network adapter is
from Mellanox and thus in theory would support the network option
\textit{\textcolor{black}{mxm}} (which uses the InfiniBand library
provided by Mellanox) for UPC, the driver version present on the system
is too old to work with UPC. We therefore use \textit{\textcolor{black}{ibv}}
(InfiniBand verbs which is a generic library used to access the InfiniBand
hardware) as the network type. The results are shown in Figure \ref{fig:Leo3}
and Table \ref{tab:Leo3}. \textit{ }

\begin{table}
\begin{centering}
\small%
\begin{tabular}{r|r@{\extracolsep{0pt}.}lr@{\extracolsep{0pt}.}l|r@{\extracolsep{0pt}.}lr@{\extracolsep{0pt}.}lr@{\extracolsep{0pt}.}lr@{\extracolsep{0pt}.}lr@{\extracolsep{0pt}.}l}
\multicolumn{15}{c}{LEO3, strong scaling}\tabularnewline
 & \multicolumn{4}{c|}{MPI} & \multicolumn{10}{c}{UPC}\tabularnewline
threads & \multicolumn{2}{c}{row} & \multicolumn{2}{c|}{patch} & \multicolumn{2}{c}{naive} & \multicolumn{2}{c}{pointer} & \multicolumn{2}{c}{barrier} & \multicolumn{2}{c}{halo} & \multicolumn{2}{c}{patch}\tabularnewline
\hline 
1 & 982&8 \phantom{00}(1.0) & 977&9 \phantom{00}(1.0) & 921&4 \phantom{0}(1.0) & 897&0 \phantom{0}(1.0) & 903&0 \phantom{0}(1.0) & \textcolor{black}{930}&\textcolor{black}{0 }\phantom{0}\textcolor{black}{(1.0)} & 943&2 \phantom{0}(1.0)\tabularnewline
4 & 287&9 \phantom{00}(3.4) & 249&1 \phantom{00}(3.9) & 253&8 \phantom{0}(3.6) & 242&6 \phantom{0}(3.7) & 243&5 \phantom{0}(3.7) & \textcolor{black}{253}&\textcolor{black}{7 }\phantom{0}\textcolor{black}{(3.7)} & 234&9 \phantom{0}(4.0)\tabularnewline
16 & 107&8 \phantom{00}(9.1) & 63&7 \phantom{0}(15.4) & 83&7 (11.0) & 81&1 (11.1) & 79&6 (11.3) & \textcolor{black}{76}&\textcolor{black}{7 (12.1)} & 72&9 (12.9)\tabularnewline
64 & 35&4 \phantom{0}(27.8) & 17&5 \phantom{0}(55.9) & 36&1 (25.5) & 34&8 (25.8) & 31&6 (28.6) & \textcolor{black}{30}&\textcolor{black}{3 (30.7)} & 27&0 (34.9)\tabularnewline
256 & 9&2 (106.8) & 6&4 (152.8) & \textcolor{black}{25}&\textcolor{black}{7 }(35.9) & 18&1 (49.6) & 14&6 (61.8) & 10&2 (91.2) & 11&7 (80.6)\tabularnewline
\multicolumn{1}{r}{} & \multicolumn{2}{c}{} & \multicolumn{2}{c}{} & \multicolumn{2}{c}{} & \multicolumn{2}{c}{} & \multicolumn{2}{c}{} & \multicolumn{2}{c}{} & \multicolumn{2}{c}{}\tabularnewline
\end{tabular}
\par\end{centering}

\begin{centering}
\small%
\begin{tabular}{rr|r@{\extracolsep{0pt}.}lr@{\extracolsep{0pt}.}l|r@{\extracolsep{0pt}.}lr@{\extracolsep{0pt}.}lr@{\extracolsep{0pt}.}lr@{\extracolsep{0pt}.}lr@{\extracolsep{0pt}.}l}
\multicolumn{16}{c}{LEO3, weak scaling}\tabularnewline
 &  & \multicolumn{4}{c|}{MPI} & \multicolumn{10}{c}{UPC}\tabularnewline
threads & grid & \multicolumn{2}{c}{row} & \multicolumn{2}{c|}{patch} & \multicolumn{2}{c}{naive} & \multicolumn{2}{c}{pointer} & \multicolumn{2}{c}{barrier} & \multicolumn{2}{c}{halo} & \multicolumn{2}{c}{patch}\tabularnewline
\hline 
1 & $64\times\phantom{0}128$  & 3&0 (1.0) & 3&0 (1.0) & 2&8 (1.0) & 2&6 (1.0) & 2&6 (1.0) & 2&5 (1.0) & 2&6 (1.0)\tabularnewline
4 & $128\times\phantom{0}256$ & 3&1 (1.0) & 3&1 (1.0) & 3&0 (1.0) & 2&6 (1.1) & 2&6 (1.0) & \textcolor{black}{2}&\textcolor{black}{6 }(1.0) & 2&6 (1.0)\tabularnewline
16 & $256\times\phantom{0}512$ & 3&6 (1.2) & 3&2 (1.1) & 3&4 (1.2) & 3&6 (1.4) & 2&9 (1.1) & \textcolor{black}{2}&\textcolor{black}{7 }(1.1) & 2&9 (1.1)\tabularnewline
64 & $512\times1024$ & 4&8 (1.6) & 3&2 (1.1) & 4&9 (1.8) & \textcolor{black}{4}&\textcolor{black}{6 }(1.8) & 3&7 (1.4) & \textcolor{black}{3}&\textcolor{black}{3 }(1.3) & 3&2 (1.2)\tabularnewline
256 & $1024\times2048$ & 6&0 (2.0) & 3&4 (1.1) & 9&5 (3.4) & \textcolor{black}{9}&\textcolor{black}{5 }(3.6) & 6&7 (2.6) & \textcolor{black}{5}&\textcolor{black}{4 }(2.1) & 3&5 (1.3)\tabularnewline
\end{tabular}
\par\end{centering}

\protect\caption{This table shows the strong and weak scaling results on the LEO3 system.
For the strong scaling analysis, we used the grid $512\times1024$.
In addition to the runtime, we also include the speedup (for the strong
scaling analysis) and the increase in runtime normalized to the single
core implementation (for the weak scaling). These quantities are shown
in parenthesis.\label{tab:Leo3}}
\end{table}

\textit{row-wise communication pattern}: On a single node, the strong
scaling analysis for all of the UPC implementations exceed the results
of the MPI version. However, as soon as we leave the node, the speedup
dramatically depends on the optimization level of the implementation.
The \textit{halo} implementation competes with the speedup of the
MPI program. Similar results are observed for the weak scaling analysis. 

\textit{patched communication pattern}: For this communication pattern,
the single node performance is similar for strong as well as weak
scaling for the MPI and the UPC version. However, the overhead for
the UPC version seems to be higher as soon as we run the simulation
on a larger number of nodes.

\subsection{Results on LEO3E}

\begin{figure}[h]
\begin{centering}
\includegraphics[scale=0.72]{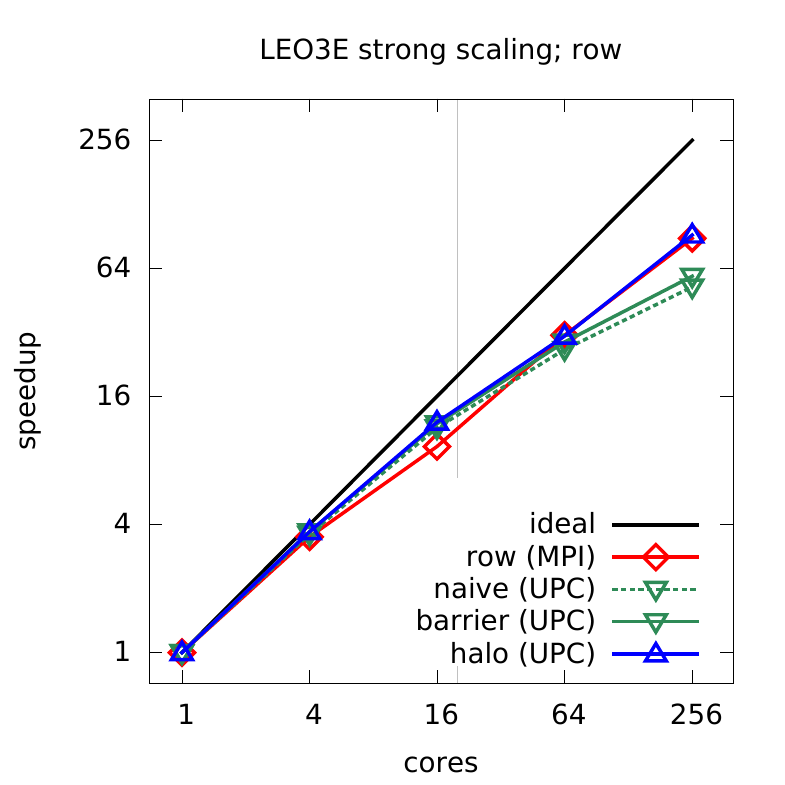}\includegraphics[scale=0.72]{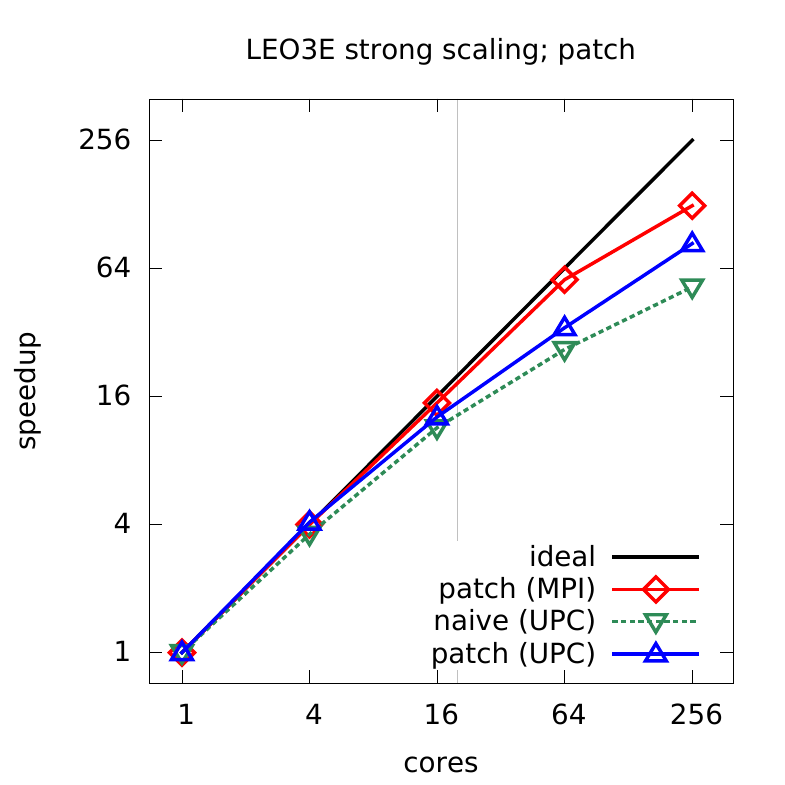}
\par\end{centering}

\begin{centering}
\includegraphics[scale=0.72]{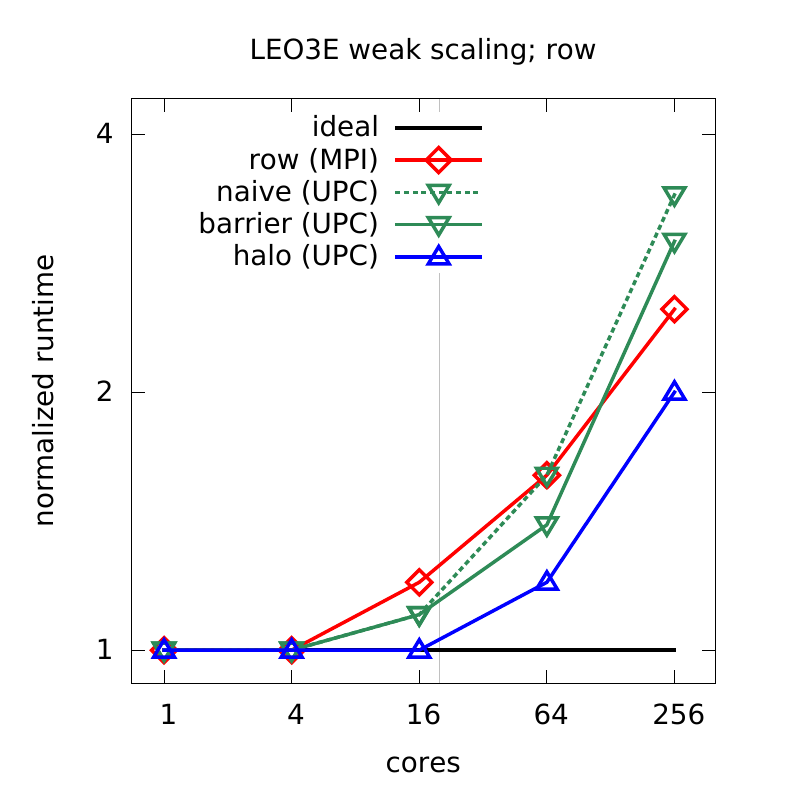}\includegraphics[scale=0.72]{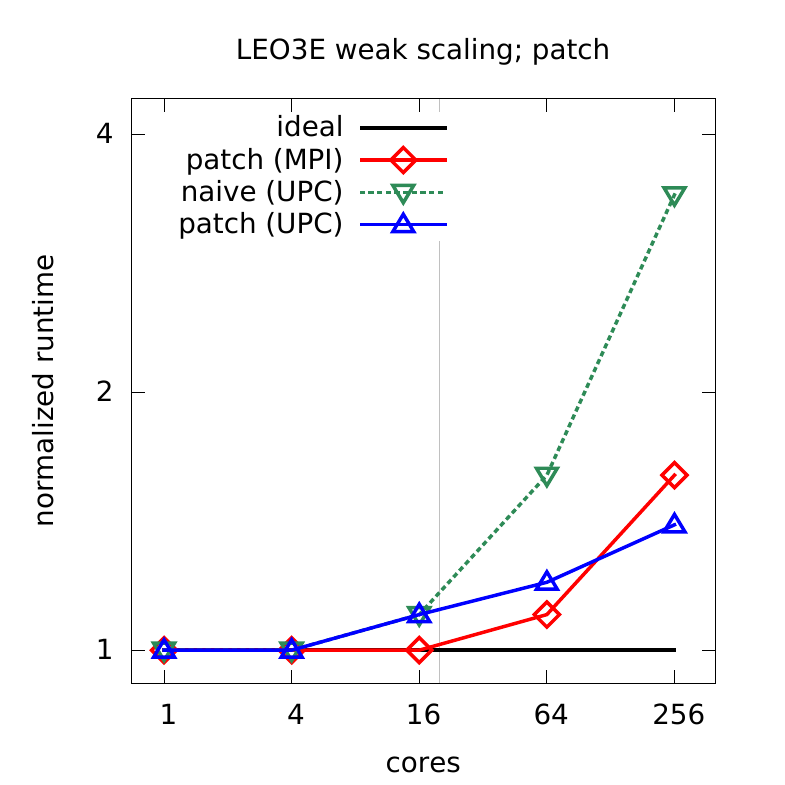}
\par\end{centering}

\protect\caption{This figure shows the scaling results on the LEO3E. The left hand
side shows from top to bottom the strong and the weak scaling results
for the \textit{row-wise} communication pattern. Due to the similar
results of the UPC \textit{row} and \textit{pointer} version, we only
include the first one in the plot. The right hand side shows from
top to bottom the strong and the weak scaling results for the \textit{patched}
communication pattern. For comparison, this column also includes the
\textit{row-wise} distributed \textit{naive} UPC version. The black
line represents the ideal scaling in all cases. The gray line represents
the number of cores where the simulation can be run on a single node.
For the LEO3E, this line is at 20 cores.\label{fig:LEO3E}}
\end{figure}

Similar to LEO3, we can not take advantage of the Mellanox network
driver for the UPC code. Therefore, \textit{\textcolor{black}{ibv}}
is used as the network type. The results for LEO3E can be found in
Figure \ref{fig:LEO3E} and Table \ref{tab:LEO3E}.

\begin{table}
\begin{centering}
\small%
\begin{tabular}{r|r@{\extracolsep{0pt}.}lr@{\extracolsep{0pt}.}l|r@{\extracolsep{0pt}.}lr@{\extracolsep{0pt}.}lr@{\extracolsep{0pt}.}lr@{\extracolsep{0pt}.}lr@{\extracolsep{0pt}.}l}
\multicolumn{15}{c}{LEO3E, strong scaling}\tabularnewline
 & \multicolumn{4}{c|}{MPI} & \multicolumn{10}{c}{UPC}\tabularnewline
threads & \multicolumn{2}{c}{row} & \multicolumn{2}{c|}{patch} & \multicolumn{2}{c}{naive} & \multicolumn{2}{c}{pointer} & \multicolumn{2}{c}{barrier} & \multicolumn{2}{c}{halo} & \multicolumn{2}{c}{patch}\tabularnewline
\hline 
1 & 849&4 \phantom{0}(1.0) & 843&9 \phantom{00}(1.0) & 616&7 \phantom{0}(1.0) & 580&9 \phantom{0}(1.0) & 599&7 \phantom{0}(1.0) & \textcolor{black}{595}&\textcolor{black}{8 }\phantom{0}(1.0) & 610&8 \phantom{0}(1.0)\tabularnewline
4 & 242&7 \phantom{0}(3.5) & 212&5 \phantom{00}(4.0) & \textcolor{black}{169}&\textcolor{black}{7} \phantom{0}(3.6) & \textcolor{black}{156}&\textcolor{black}{9} \phantom{0}(3.7) & \textcolor{black}{161}&\textcolor{black}{0} \phantom{0}(3.7) & \textcolor{black}{159}&\textcolor{black}{4 }\phantom{0}(3.7) & 148&7 \phantom{0}(4.1)\tabularnewline
16 & 91&4 \phantom{0}(9.3) & 56&5 \phantom{0}(14.9) & \textcolor{black}{54}&\textcolor{black}{3} (11.4) & 50&8 (11.4) & \textcolor{black}{50}&\textcolor{black}{3 }(11.9) & \textcolor{black}{49}&\textcolor{black}{4 }(12.1) & 47&9 (12.8)\tabularnewline
64 & 27&4 (31.0) & 14&9 \phantom{0}(56.6) & \textcolor{black}{23}&2 (26.6) & 23&4 (24.8) & \textcolor{black}{20}&\textcolor{black}{8} (28.8) & \textcolor{black}{19}&\textcolor{black}{4} (30.7) & 18&2 (33.6)\tabularnewline
256 & 9&6 (88.5) & 6&7 (126.0) & 11&8 (52.3) & 11&2 (51.9) & 10&2 (58.8) & 6&5 (91.7) & 7&3 (83.7)\tabularnewline
\multicolumn{1}{r}{} & \multicolumn{2}{c}{} & \multicolumn{2}{c}{} & \multicolumn{2}{c}{} & \multicolumn{2}{c}{} & \multicolumn{2}{c}{} & \multicolumn{2}{c}{} & \multicolumn{2}{c}{}\tabularnewline
\end{tabular}
\par\end{centering}

\begin{centering}

\par\end{centering}

\begin{centering}
\small%
\begin{tabular}{rr|r@{\extracolsep{0pt}.}lr@{\extracolsep{0pt}.}l|r@{\extracolsep{0pt}.}lr@{\extracolsep{0pt}.}lr@{\extracolsep{0pt}.}lr@{\extracolsep{0pt}.}lr@{\extracolsep{0pt}.}l}
\multicolumn{16}{c}{LEO3E, weak scaling}\tabularnewline
 &  & \multicolumn{4}{c|}{MPI} & \multicolumn{10}{c}{UPC}\tabularnewline
threads & grid & \multicolumn{2}{c}{row} & \multicolumn{2}{c|}{patch} & \multicolumn{2}{c}{naive} & \multicolumn{2}{c}{pointer} & \multicolumn{2}{c}{barrier} & \multicolumn{2}{c}{halo} & \multicolumn{2}{c}{patch}\tabularnewline
\hline 
1 & $64\times\phantom{0}128$  & 2&6 (1.0) & 2&5 (1.0) & 1&8 (1.0) & 1&7 (1.0) & 1&7 (1.0) & \textcolor{black}{1}&\textcolor{black}{7 }(1.0) & 1&7 (1.0)\tabularnewline
4 & $128\times\phantom{0}256$ & 2&6 (1.0) & 2&6 (1.0) & 1&8 (1.0) & 1&7 (1.0) & 1&7 (1.0) & \textcolor{black}{1}&\textcolor{black}{7 }(1.0) & 1&7 (1.0)\tabularnewline
16 & $256\times\phantom{0}512$ & 3&0 (1.2) & 2&6 (1.0) & 2&0 (1.1) & 1&9 (1.1) & 1&9 (1.1) & \textcolor{black}{1}&\textcolor{black}{7 }(1.0) & 1&9 (1.1)\tabularnewline
64 & $512\times1024$ & 4&1 (1.6) & 2&7 (1.1) & 2&9 (1.6) & 2&8 (1.6) & 2&4 (1.4) & \textcolor{black}{2}&\textcolor{black}{0 }(1.2) & 2&1 (1.2)\tabularnewline
256 & $1024\times2048$ & 6&5 (2.5) & 4&1 (1.6) & 6&2 (3.4) & 6&0 (3.5) & 5&1 (3.0) & 3&4 (2.0) & 2&4 (1.4)\tabularnewline
\end{tabular}
\par\end{centering}

\protect\caption{This table shows the strong and weak scaling results on the LEO3E
system. For the strong scaling analysis, we used the grid $512\times1024$.
In addition to the runtime, we also include the speedup (for the strong
scaling analysis) and the increase in runtime normalized to the single
core implementation (for the weak scaling). These quantities are shown
in parenthesis.\label{tab:LEO3E}}
\end{table}

The processors on this hardware are newer and thus faster than on
LEO3. This can be seen in the significantly shorter runtime of the
sequential code (see, e.g., Table \ref{tab:LEO3E}). 

\textit{row-wise communication pattern}: On a single node, the different
optimization stages of the UPC code are similar and compete with the
MPI version. For a larger number of nodes, the \textit{halo} version
scales better compared to the MPI program. This is valid for both
the strong as well as the weak scaling analysis.

\textit{patched communication pattern}: The speedup for the strong
scaling analysis of the MPI version is better than the UPC program.
However, for the weak scaling analysis, UPC scales better compared
to the MPI version.

\subsection{Results on VSC2}

\begin{figure}[h]
\begin{centering}
\includegraphics[scale=0.72]{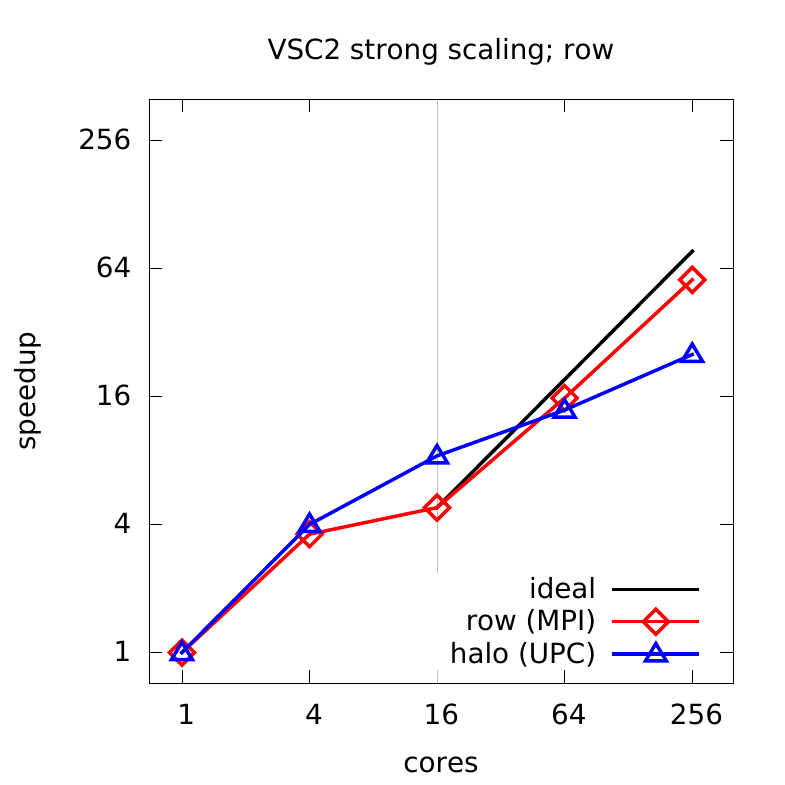}\includegraphics[scale=0.72]{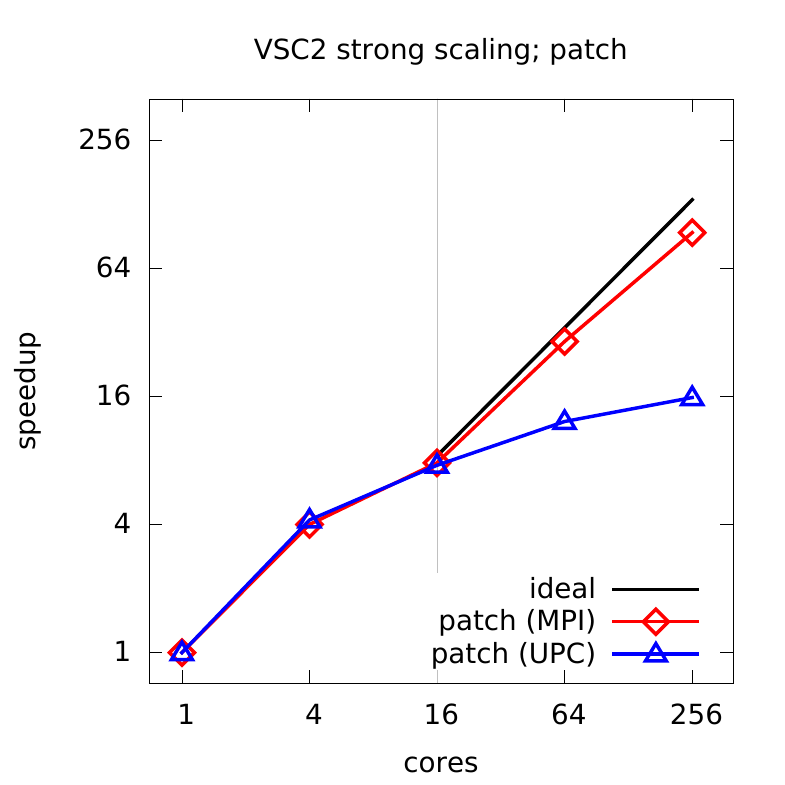}
\par\end{centering}

\begin{centering}
\includegraphics[scale=0.72]{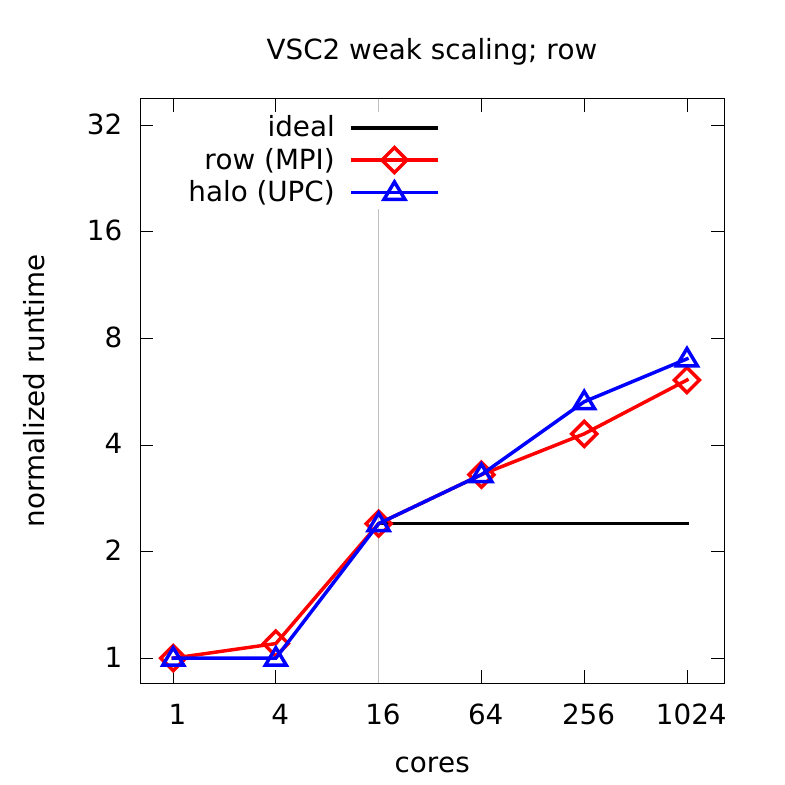}\includegraphics[scale=0.72]{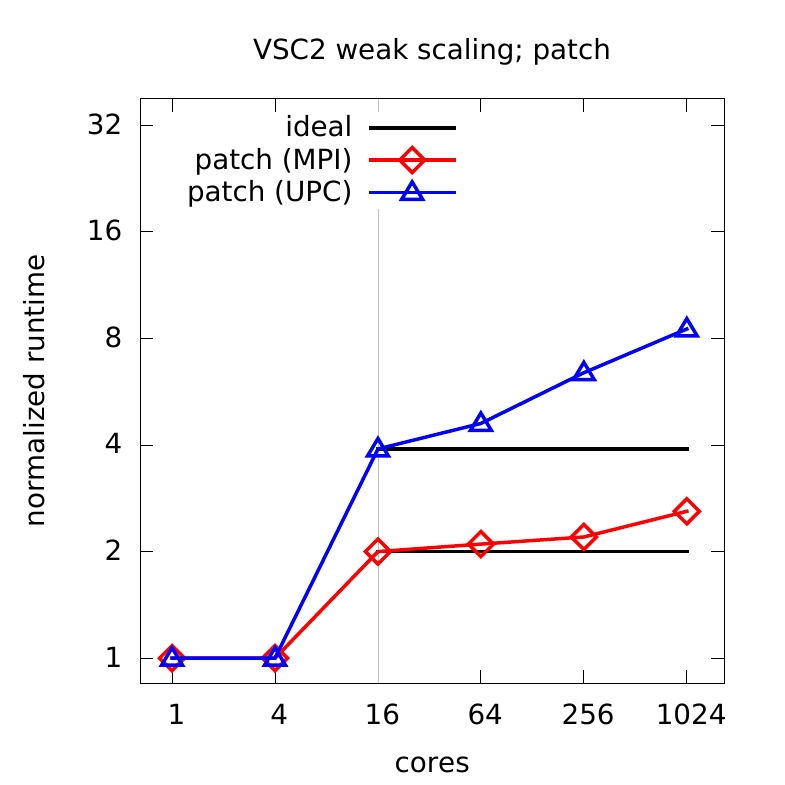} 
\par\end{centering}

\protect\caption{This figure shows the scaling results on the VSC2. The left hand side
shows from top to bottom the strong and the weak scaling results for
the \textit{row-wise} communication pattern. Due to the significant
run time increase for the \textit{naive}, \textit{pointer}, and \textit{barrier}
UPC runs, we only compare the \textit{halo} implementation. The right
hand side shows from top to bottom the strong and the weak scaling
results for the \textit{patched} communication pattern. Since the
architecture of the AMD CPU only includes a single floating point
unit for every two cores, perfect scaling on one node is not expected
and not achieved. In that respect, we only included the ideal scaling
line in black when the simulation leaves one node. The gray line represents
the number of cores where the simulation can be run on a single node.
For the VSC2, this line is at 16 cores.\label{fig:VSC2}}
\end{figure}

VSC2 uses a Mellanox adapter and we are able to make use of the \textit{mxm}-network
type. For this system, we experience a significant performance issue
brought about by the synchronization barriers in our code. Due to
the significant run time increase for the \textit{naive}, \textit{pointer},
and \textit{barrier} UPC runs, we only compare the \textit{halo} implementation
with MPI (Figure \ref{fig:VSC2} and Table \ref{tab:VSC2}). Since
the VSC2 system is larger than LEO3 and LEO3E, we perform a weak scaling
analysis for up to 1024 cores. This can not be done for strong scaling,
since due to the data distribution of the UPC code, we have not enough
data points for the problem size considered.

\begin{table}
\begin{centering}
\small%
\begin{tabular}{r|r@{\extracolsep{0pt}.}lr@{\extracolsep{0pt}.}l|r@{\extracolsep{0pt}.}lr@{\extracolsep{0pt}.}l}
\multicolumn{9}{c}{VSC2, strong scaling}\tabularnewline
 & \multicolumn{4}{c|}{MPI} & \multicolumn{4}{c}{UPC}\tabularnewline
threads & \multicolumn{2}{c}{row} & \multicolumn{2}{c|}{patch} & \multicolumn{2}{c}{halo} & \multicolumn{2}{c}{patch}\tabularnewline
\hline 
1 & 1615&6 \phantom{0}(1.0\phantom{*}) & 1535&7 \phantom{0}(1.0\phantom{*}) & \textcolor{black}{1649}&\textcolor{black}{9} (1.0\phantom{*}) & 1653&2 (1.0\phantom{*})\tabularnewline
4 & 451&4 \phantom{0}(3.6\phantom{*}) & 380&6 \phantom{0}(4.0\phantom{*}) & \textcolor{black}{415}&\textcolor{black}{2 }(4.0\phantom{*}) & 391&9 (4.2\phantom{*})\tabularnewline
16 & 337&2 \phantom{0}(4.8\phantom{*}) & 196&6 \phantom{0}(7.8\phantom{*}) & \textcolor{black}{197}&\textcolor{black}{5 }(8.4\phantom{*}) & 217&0 (7.6\phantom{*})\tabularnewline
\hline 
64 & 102&4 \phantom{0}(3.3{*}) & 52&9 \phantom{0}(3.8{*}) & \textcolor{black}{119}&\textcolor{black}{9 }(1.6{*}) & 135&6 (1.6{*})\tabularnewline
256 & 28&6 (11.8{*}) & 16&3 (12.1{*}) & \textcolor{black}{65}&\textcolor{black}{4 }(3.0{*}) & 104&9 (2.1{*})\tabularnewline
\multicolumn{1}{r}{} & \multicolumn{2}{c}{} & \multicolumn{2}{c}{} & \multicolumn{2}{c}{} & \multicolumn{2}{c}{}\tabularnewline
\end{tabular}
\par\end{centering}

\begin{centering}

\par\end{centering}

\begin{centering}
\small%
\begin{tabular}{rr|r@{\extracolsep{0pt}.}lr@{\extracolsep{0pt}.}l|r@{\extracolsep{0pt}.}lr@{\extracolsep{0pt}.}l}
\multicolumn{10}{c}{VSC2, weak scaling}\tabularnewline
 &  & \multicolumn{4}{c|}{MPI} & \multicolumn{4}{c}{UPC}\tabularnewline
threads & grid & \multicolumn{2}{c}{row} & \multicolumn{2}{c|}{patch} & \multicolumn{2}{c}{halo} & \multicolumn{2}{c}{patch}\tabularnewline
\hline 
1 & $64\times\phantom{0}128$  & 4&5 (1.0\phantom{*}) & 4&5 (1.0\phantom{*}) & \textcolor{black}{4}&\textcolor{black}{4 }(1.0\phantom{*}) & 4&4 (1.0\phantom{*})\tabularnewline
4 & $128\times\phantom{0}256$ & 4&8 (1.1\phantom{*}) & 4&6 (1.0\phantom{*}) & \textcolor{black}{4}&\textcolor{black}{3 }(1.0\phantom{*}) & 4&2 (1.0\phantom{*})\tabularnewline
16 & $256\times\phantom{0}512$ & 11&0 (2.4\phantom{*}) & 9&2 (2.0\phantom{*}) & \textcolor{black}{10}&\textcolor{black}{5 }(2.4\phantom{*}) & 17&0 (3.9\phantom{*})\tabularnewline
\hline 
64 & $512\times1024$ & 14&9 (1.4{*}) & 9&6 (1.0{*}) & \textcolor{black}{14}&\textcolor{black}{9 }(1.4{*}) & 20&1 (1.2{*})\tabularnewline
256 & $1024\times2048$ & 19&3 (1.8{*}) & 9&8 (1.1{*}) & \textcolor{black}{23}&\textcolor{black}{2 }(2.2{*}) & 28&1 (1.7{*})\tabularnewline
1024 & $2048\times4096$ & 27&5 (2.5{*}) & 11&6 (1.3{*}) & 30&9 (2.9{*}) & 37&3 (2.2{*})\tabularnewline
\end{tabular}
\par\end{centering}

\protect\caption{This table shows the strong and weak scaling results on the VSC2 system.
For the strong scaling analysis, we used the grid $512\times1024$.
In addition to the runtime, we also include the speedup (for the strong
scaling analysis) and the increase in runtime normalized to the single
core implementation (for the weak scaling). These quantities are shown
in parenthesis. The numbers denoted with {*} are not based on one
core but on 16 cores due to the hardware properties described in the
text.\label{tab:VSC2}}
\end{table}

Due to the architecture of the AMD CPU, which includes only a single
floating point unit for every two cores, we only observe a speedup
of 8 on a single node. This is an inherent limitation of the CPU and
can be observed for both the MPI as well as the UPC implementation.

\textit{row-wise communication pattern}: We observe that the single
node performance of the UPC \textit{halo} implementation outperforms
the MPI version. However, as soon as we use a higher number of nodes,
the performance is disappointing for UPC. This is a result of the
fact that a large portion of the run time is spend in the two calls
(per time step) to \textit{upc\_barrier} (see Table \ref{tab:VSC2}).

\textit{patched communication pattern}: Similar to the row pattern
implementations, the single node performance of the MPI and the UPC
implementation is comparable, however, the more cores are used, the
more time is spend at the \textit{upc\_barrier} calls that significantly
increases the run time of UPC. However, it should be noted that the
scaling behavior of the MPI code is also far from ideal on this system.

\subsection{Results on VSC3}

VSC3 has an InfiniBand network adapter from Intel. Since Berkeley
UPC does not include native support for this hardware, we use\textit{
ibv} as the network type. VSC3 is the largest system we have at our
disposal. We therefore perform the weak scaling analysis for up to
1024 cores. 

\begin{figure}[h]
\begin{centering}
\includegraphics[scale=0.72]{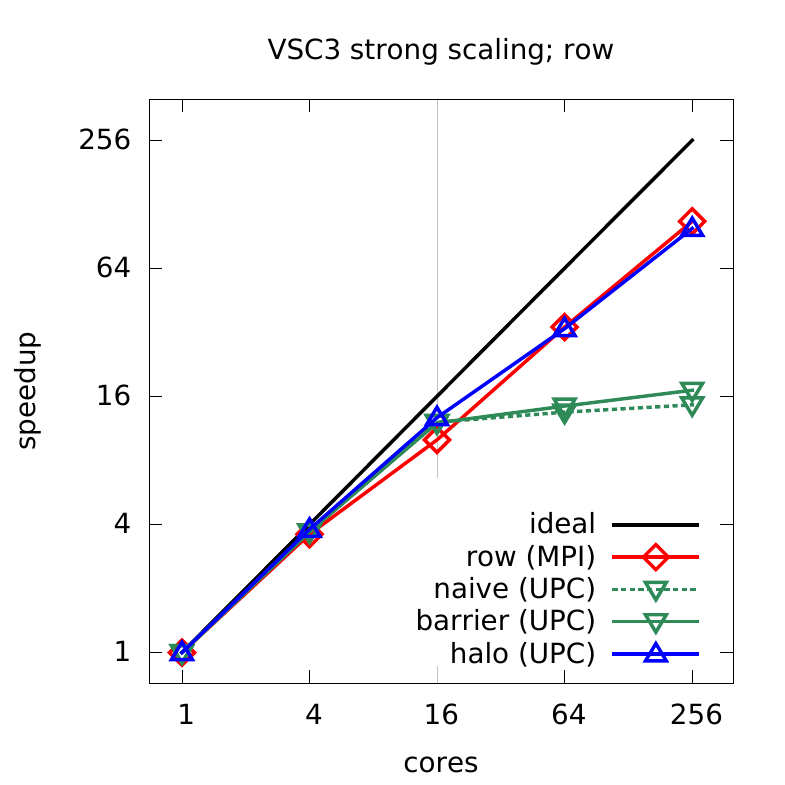}\includegraphics[scale=0.72]{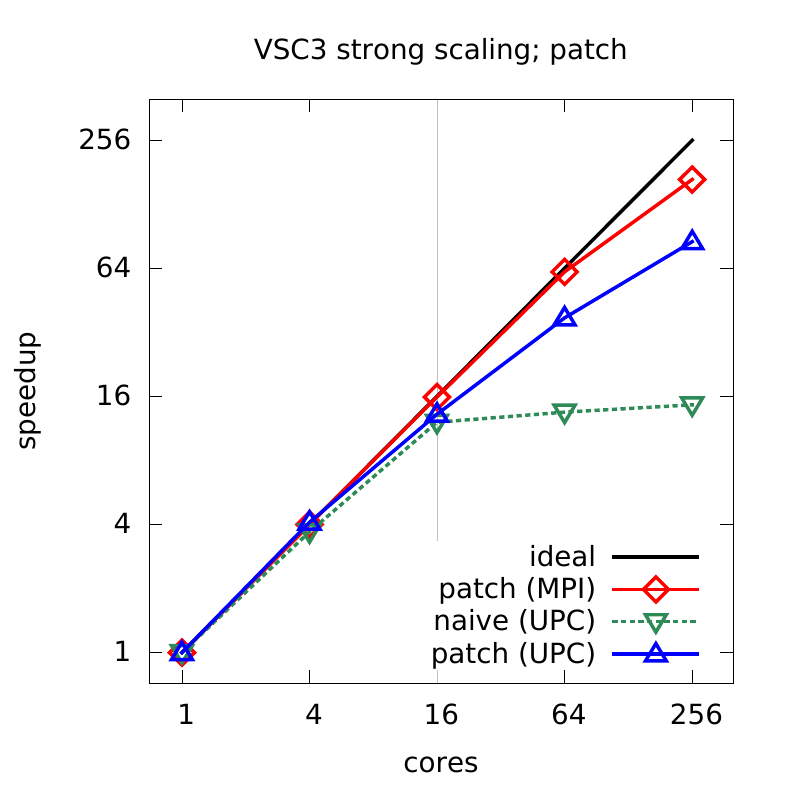}
\par\end{centering}

\begin{centering}
\includegraphics[scale=0.72]{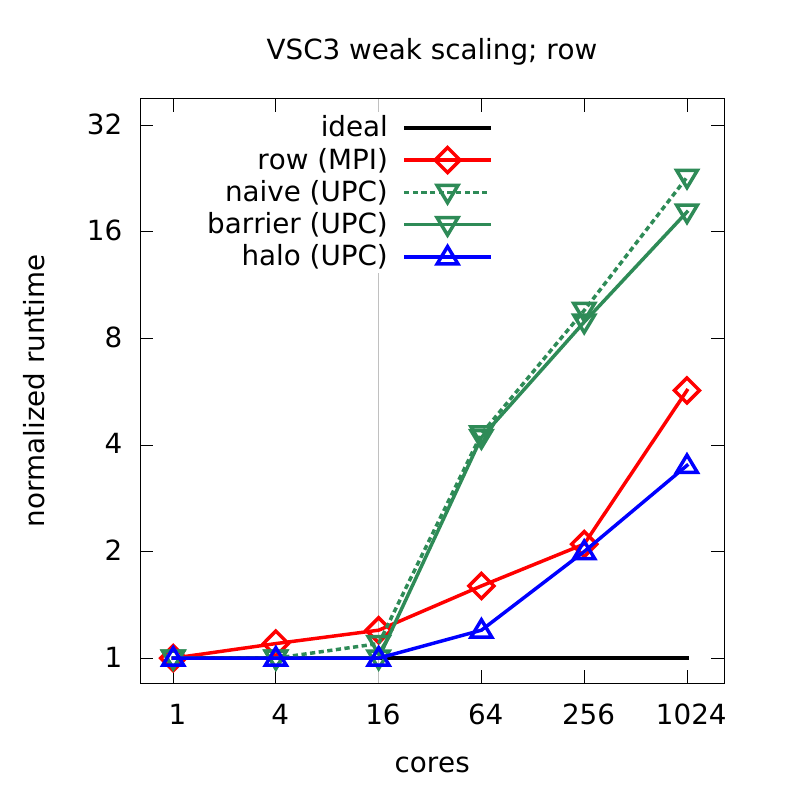}\includegraphics[scale=0.72]{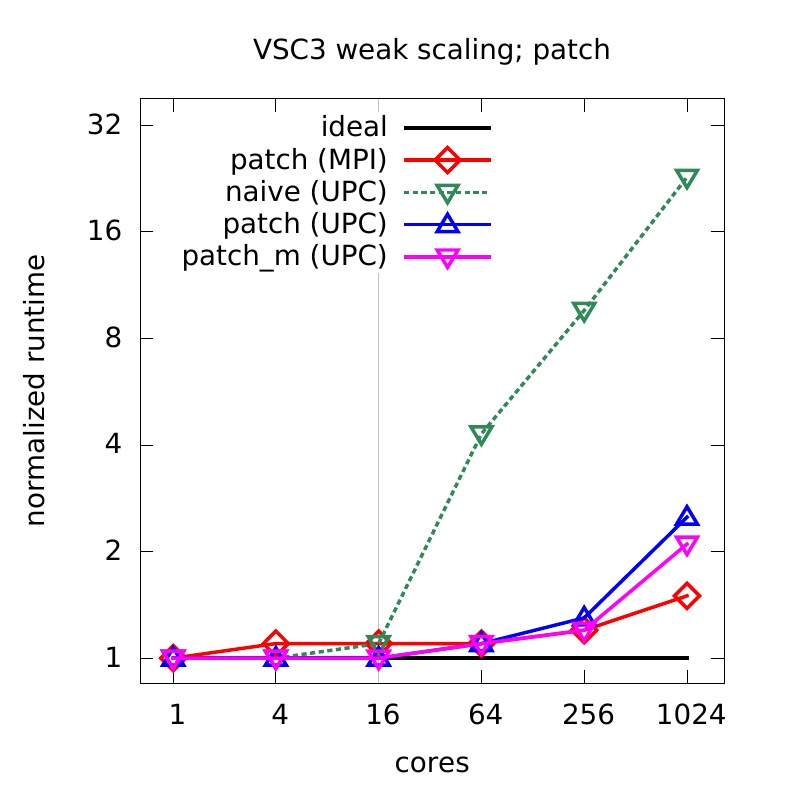} 
\par\end{centering}

\protect\caption{This figure shows the scaling results on the VSC3. The left hand side
shows from top to bottom the strong and the weak scaling results for
the \textit{row-wise} communication pattern. Due to the similar results
of the UPC \textit{row} and \textit{pointer} version, we only include
the first one in the plot. The right hand side shows from top to bottom
the strong and the weak scaling results for the \textit{patched} communication
pattern. For comparison, this column also includes the \textit{row-wise}
distributed \textit{naive} UPC version and an additional modified
UPC \textit{patch} version \textit{patch\_m} (\textit{patch\_mod}
in Table \ref{tab:VSC3}). In this version, more data is communicated
in less communication steps. The black line represents the ideal scaling
in all cases. The gray line represents the number of cores where the
simulation can be run on a single node. For the VSC3, this line is
at 16 cores.\label{fig:VSC3}}
\end{figure}

However, for problems with 1024 or more cores, by default, the UPC
run time opens more InfiniBand connections than the hardware supports.
In principle, to address this issue, XRC and SRQ were developed. Unfortunately,
these technologies are not supported on VSC3. Thus, we have to bundle
the network connections manually by reducing the number of processes
on a single node. In UPC, this is accomplished by setting the pthread
option on the command line. Increasing the number of pthreads also
leads to an increase of the run time. We therefore only use the smallest
number of pthreads possible. I.e., for $1024$ threads, we use $pthreads=2$.
The detailed results can be found in Table \ref{tab:VSC3} and a visual
representation of the data is shown in Figure \ref{fig:VSC3}.

\begin{table}
\begin{centering}

\par\end{centering}

\begin{centering}
\small%
\begin{tabular}{c|r@{\extracolsep{0pt}.}lr@{\extracolsep{0pt}.}l}
\multicolumn{5}{c}{VSC3, strong scaling, MPI}\tabularnewline
threads & \multicolumn{2}{c}{row} & \multicolumn{2}{c}{patch}\tabularnewline
\hline 
1 & 532&2 \phantom{00}(1.0) & \textcolor{black}{535}&\textcolor{black}{8 }\phantom{00}\textcolor{black}{(1.0)}\tabularnewline
4 & 149&5 \phantom{00}(3.6) & \textcolor{black}{132}&\textcolor{black}{9 }\phantom{00}\textcolor{black}{(4.0)}\tabularnewline
16 & 53&2 \phantom{0}(10.0) & \textcolor{black}{33}&\textcolor{black}{7 }\phantom{0}\textcolor{black}{(15.9)}\tabularnewline
64 & 15&7 \phantom{0}(33.9) & \textcolor{black}{8}&\textcolor{black}{7 }\phantom{0}\textcolor{black}{(61.6)}\tabularnewline
256 & 5&0 (106.4) & \textcolor{black}{3}&\textcolor{black}{2 (167.4)}\tabularnewline
\multicolumn{1}{c}{} & \multicolumn{2}{c}{} & \multicolumn{2}{c}{}\tabularnewline
\multicolumn{1}{c}{} & \multicolumn{2}{c}{} & \multicolumn{2}{c}{}\tabularnewline
\end{tabular}\qquad \small%
\begin{tabular}{c|cr@{\extracolsep{0pt}.}lr@{\extracolsep{0pt}.}l}
\multicolumn{6}{c}{VSC3, weak scaling, MPI}\tabularnewline
threads & grid & \multicolumn{2}{c}{row} & \multicolumn{2}{c}{patch}\tabularnewline
\hline 
1 & $64\times\phantom{0}128$  & 1&5 (1.0) & \textcolor{black}{1}&\textcolor{black}{5 (1.0)}\tabularnewline
4 & $128\times\phantom{0}256$ & 1&6 (1.1) & 1&6 (1.1)\tabularnewline
16 & $256\times\phantom{0}512$ & 1&8 (1.2) & \textcolor{black}{1}&\textcolor{black}{6 (1.1)}\tabularnewline
64 & $512\times1024$ & 2&4 (1.6) & 1&6 (1.1)\tabularnewline
256 & $1024\times2048$ & 3&2 (2.1) & \textcolor{black}{1}&\textcolor{black}{8 (1.2)}\tabularnewline
1024 & $2048\times4096$ & 8&8 (5.7) & \textcolor{black}{2}&\textcolor{black}{3 (1.5)}\tabularnewline
\multicolumn{1}{c}{} &  & \multicolumn{2}{c}{} & \multicolumn{2}{c}{}\tabularnewline
\end{tabular}\\
\small%
\begin{tabular}{c|r@{\extracolsep{0pt}.}lr@{\extracolsep{0pt}.}lr@{\extracolsep{0pt}.}lr@{\extracolsep{0pt}.}lr@{\extracolsep{0pt}.}lr@{\extracolsep{0pt}.}l}
\multicolumn{13}{c}{VSC3, strong scaling, UPC}\tabularnewline
threads & \multicolumn{2}{c}{naive} & \multicolumn{2}{c}{pointer} & \multicolumn{2}{c}{barrier} & \multicolumn{2}{c}{halo} & \multicolumn{2}{c}{patch} & \multicolumn{2}{c}{patch\_mod}\tabularnewline
\hline 
1 & 516&5 \phantom{0}\textcolor{black}{(1.0)} & 459&1 \phantom{0}\textcolor{black}{(1.0)} & 478&1 \phantom{0}\textcolor{black}{(1.0)} & \textcolor{black}{482}&\textcolor{black}{2 }\phantom{0}\textcolor{black}{(1.0)} & 478&3 \phantom{0}(1.0) & 478&0 \phantom{0}(1.0)\tabularnewline
4 & 141&2 \phantom{0}\textcolor{black}{(3.7)} & 123&8 \phantom{0}\textcolor{black}{(3.7)} & 128&2 \phantom{0}\textcolor{black}{(3.7)} & \textcolor{black}{125}&\textcolor{black}{9 }\phantom{0}\textcolor{black}{(3.8)} & 116&8 \phantom{0}(4.1) & 117&9 \phantom{0}(4.1)\tabularnewline
16 & 42&6 \textcolor{black}{(12.1)} & 37&8 \textcolor{black}{(12.1)} & 39&8 \textcolor{black}{(12.0)} & \textcolor{black}{38}&\textcolor{black}{0 (12.7)} & 36&3 (13.2) & 36&1 (13.2)\tabularnewline
64 & 38&3 \textcolor{black}{(13.5)} & 37&0 \textcolor{black}{(12.4)} & 33&3 \textcolor{black}{(14.4)} & \textcolor{black}{14}&\textcolor{black}{5 (33.3)} & 12&8 (37.4) & 12&7 (37.6)\tabularnewline
256 & 35&3 \textcolor{black}{(14.6)} & 34&2 \textcolor{black}{(13.4)} & 28&0 \textcolor{black}{(17.1)} & \textcolor{black}{4}&\textcolor{black}{9 (98.4)} & 5&6 (85.4) & 5&5 (86.9)\tabularnewline
\multicolumn{1}{c}{} & \multicolumn{2}{c}{} & \multicolumn{2}{c}{} & \multicolumn{2}{c}{} & \multicolumn{2}{c}{} & \multicolumn{2}{c}{} & \multicolumn{2}{c}{}\tabularnewline
\end{tabular}\\
\small%
\begin{tabular}{c|r@{\extracolsep{0pt}.}lr@{\extracolsep{0pt}.}lr@{\extracolsep{0pt}.}lr@{\extracolsep{0pt}.}lr@{\extracolsep{0pt}.}lr@{\extracolsep{0pt}.}l}
\multicolumn{13}{c}{VSC3, weak scaling, UPC}\tabularnewline
threads & \multicolumn{2}{c}{naive} & \multicolumn{2}{c}{pointer} & \multicolumn{2}{c}{barrier} & \multicolumn{2}{c}{halo} & \multicolumn{2}{c}{patch} & \multicolumn{2}{c}{patch\_mod}\tabularnewline
\hline 
1 & 1&5 \phantom{0}\phantom{0}\textcolor{black}{(1.0)} & 1&3 \phantom{0}\phantom{0}\textcolor{black}{(1.0)} & 1&3 \phantom{0}\phantom{0}\textcolor{black}{(1.0)} & \textcolor{black}{1}&\textcolor{black}{3}\phantom{*}\textcolor{black}{{} (1.0)} & 1&3\phantom{*} (1.0) & 1&3\phantom{*} (1.0)\tabularnewline
4 & 1&5 \phantom{0}\phantom{0}\textcolor{black}{(1.0)} & 1&3 \phantom{0}\phantom{0}\textcolor{black}{(1.0)} & 1&3 \phantom{0}\phantom{0}\textcolor{black}{(1.0)} & \textcolor{black}{1}&\textcolor{black}{3}\phantom{*}\textcolor{black}{{} (1.0)} & 1&3\phantom{*} (1.0) & 1&3\phantom{*} (1.0)\tabularnewline
16 & 1&6 \phantom{0}\phantom{0}\textcolor{black}{(1.1)} & 1&4 \phantom{0}\phantom{0}\textcolor{black}{(1.1)} & 1&3 \phantom{0}\phantom{0}\textcolor{black}{(1.0)} & \textcolor{black}{1}&\textcolor{black}{3}\phantom{*}\textcolor{black}{{} (1.0)} & 1&3\phantom{*} (1.0) & 1&3\phantom{*} (1.0)\tabularnewline
64 & 6&4 \phantom{0}\phantom{0}\textcolor{black}{(4.3)} & 6&1 \phantom{0}\phantom{0}\textcolor{black}{(4.7)} & 5&6 \phantom{0}\phantom{0}\textcolor{black}{(4.2)} & \textcolor{black}{1}&\textcolor{black}{6}\phantom{*}\textcolor{black}{{} (1.2)} & 1&4\phantom{*} (1.1) & 1&4\phantom{*} (1.1)\tabularnewline
256 & 14&3 \phantom{0}\phantom{0}\textcolor{black}{(9.6)} & 14&6 \phantom{0}\textcolor{black}{(11.2)} & 11&8 \phantom{0}\phantom{0}\textcolor{black}{(8.9)} & \textcolor{black}{2}&\textcolor{black}{6}\phantom{*}\textcolor{black}{{} (2.0)} & 1&7\phantom{*} (1.3) & 1&6\phantom{*} (1.2)\tabularnewline
1024 & 34&2{*} (22.8) & 33&8{*} (26.0) & 23&7{*} (18.2) & \textcolor{black}{4}&\textcolor{black}{5{*} (3.5)} & 3&3{*} (2.5) & 2&7{*} (2.1)\tabularnewline
\end{tabular}
\par\end{centering}

\protect\caption{This table shows the strong and weak scaling results on the VSC3 system.
For the strong scaling analysis, we used the grid $512\times1024$.
In addition to the runtime, we also include the speedup (for the strong
scaling analysis) and the increase in runtime normalized to the single
core implementation (for the weak scaling). These quantities are shown
in parenthesis. The numbers marked with a {*} are simulated by using
2 pthreads. This issue is discussed further within the text.\label{tab:VSC3}}
\end{table}

In this simulation, we have the newest as well as largest cluster
within this paper at our disposal. This already affects the run time
for one core, which is nearly half of the run time on the LEO3 system
for all of the runs investigated.

\textit{row-wise communication pattern}: Similar to the other hardware,
all UPC optimization stages perform similarly on the same node. However,
for a larger number of nodes, only the UPC \textit{halo} version is
able to compete with the MPI implementation. The strong scaling analysis
is comparable for both implementations. For $1024$ cores, the weak
scaling results for the \textit{halo} UPC implementation significantly
outperforms the MPI implementation.

\textit{patched communication pattern}: The MPI version shows better
scaling results than the UPC version. Since on this hardware, a significant
amount of time is spend in the barriers, we also include the results
for a modified \textit{patch} version \textit{patch\_mod}: In this
version, we communicate the data both in the $x$- as well as the
$y$-direction before the splitting. This means that in each time
step, we only communicate once, but twice as many data. This enables
us to get rid of two barriers that are necessary to avoid race conditions
in the \textit{patch} implementation. Note that this is a first order
approximation, even though the boundary data in the $y$-direction
is not updated with the output from the splitting step in the $x$-direction,
but from the data of the previous time step. 

As we can see, the runtime for both patched versions only differs,
when we reach a large number of nodes. On LEO3, LEO3E and VSC2 we
observed no performance difference for these two versions. For this
reason, we only include this additional optimization for  VSC3. We
can see that especially for more than a thousand cores, the overhead
of the barriers have a significant impact. However, for this communication
pattern, UPC can still not compete with MPI on the VSC3.

\subsection{Results on the Intel Xeon Phi}

\begin{figure}[h]
\begin{centering}
\includegraphics[scale=0.72]{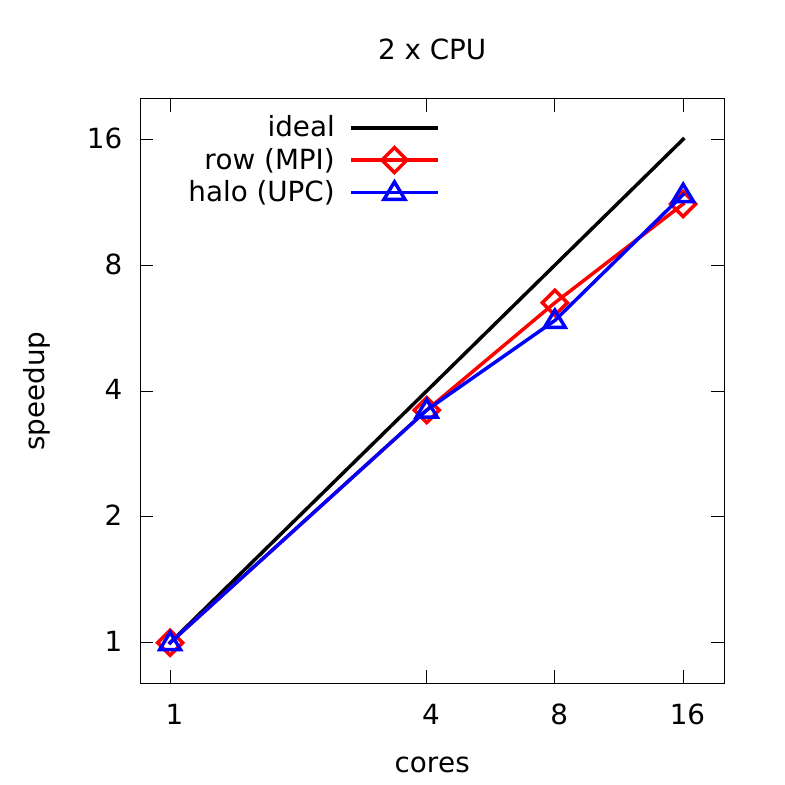}\includegraphics[scale=0.72]{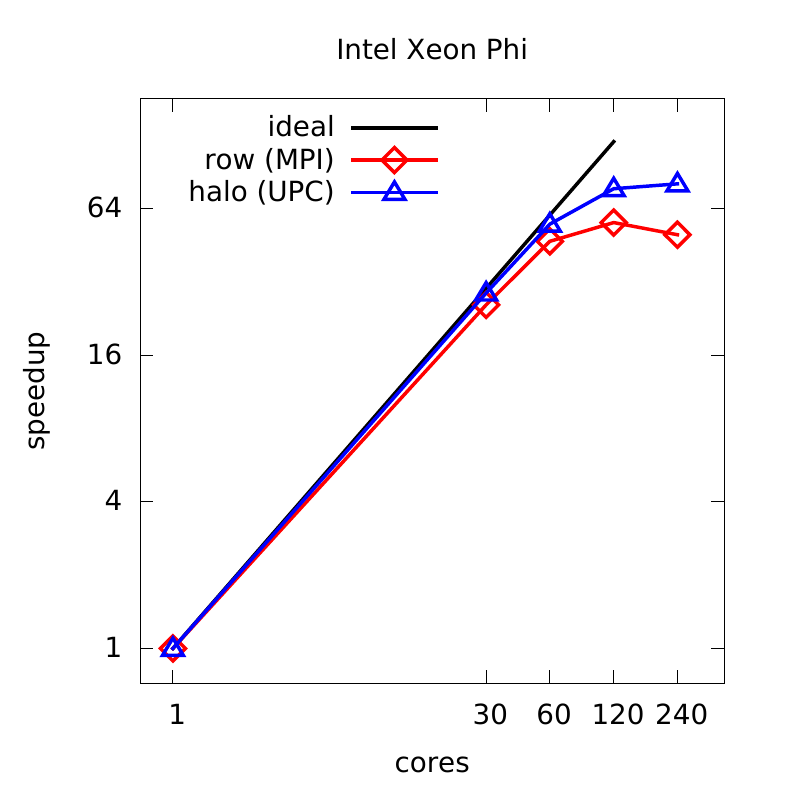}
\par\end{centering}

\protect\caption{This figure shows the strong scaling results for the Intel Xeon Phi.
The left hand side shows the scaling behavior of the two CPUs (Intel
Xeon E5-2630 v3) present in that system. The right hand side shows
the scaling on the Intel Xeon Phi 7120 for up to 240 threads. The
black line represents the ideal scaling in both cases.\label{fig:xeonphi}}
\end{figure}

Note that the Intel Xeon Phi usually benefits from additional vectorization
optimizations. However, in this paper we use our most optimized UPC
run as well as our MPI run without any change. The Intel Xeon Phi
has 60 cores and 4 hyperthreads at each core that we could use. Note,
however, that even with hyperthreading, we do not expect a linear
increase after 60 cores. We only use the \textit{row} implementation
for MPI and the UPC \textit{halo} version in this section. In Figure
\ref{fig:xeonphi} and Table \ref{tab:xeonphi}, we show the results
for strong scaling for both CPUs on a single node and for the Intel
Xeon Phi. 

\begin{table}
\begin{centering}
\begin{tabular}{r|r@{\extracolsep{0pt}.}lr@{\extracolsep{0pt}.}l}
\multicolumn{5}{c}{2$\times$CPU}\tabularnewline
threads & \multicolumn{2}{c}{MPI: row} & \multicolumn{2}{c}{UPC: halo}\tabularnewline
\hline 
1 & 3&58 \phantom{0}(1.0) & 3&08 \phantom{0}(1.0)\tabularnewline
4 & 0&99 \phantom{0}(3.6) & 0&86 \phantom{0}(3.6)\tabularnewline
8 & 0&55 \phantom{0}(6.5) & 0&52 \phantom{0}(5.9)\tabularnewline
16 & 0&32 (11.2) & 0&26 (11.8)\tabularnewline
\multicolumn{1}{r}{} & \multicolumn{2}{c}{} & \multicolumn{2}{c}{}\tabularnewline
\end{tabular}$\qquad$%
\begin{tabular}{r|r@{\extracolsep{0pt}.}lr@{\extracolsep{0pt}.}l}
\multicolumn{5}{c}{Intel Xeon Phi}\tabularnewline
threads & \multicolumn{2}{c}{MPI: row} & \multicolumn{2}{c}{UPC: halo}\tabularnewline
\hline 
1 & 45&83 \phantom{0}(1.0) & 33&06 \phantom{0}(1.0)\tabularnewline
30 & 1&78 (25.7) & 1&15 (28.7)\tabularnewline
60 & 0&98 (46.8) & 0&60 (55.1)\tabularnewline
120 & 0&82 (55.9) & 0&43 (76.9)\tabularnewline
240 & 0&92 (49.8) & 0&41 (80.6)\tabularnewline
\end{tabular}
\par\end{centering}

\protect\caption{This table shows the strong scaling results with the grid $512\times1024$
on the Intel Xeon Phi system. The left analysis is performed on the
two E5-2630v3 CPUs within that system. The left analysis is performed
on the Intel Xeon Phi 7120. We also include the speedup in parenthesis.
The grid size in $y$-direction is a multiple of $60$. The final
integration time is $T=5\cdot10^{-5}$.\label{tab:xeonphi}}
\end{table}

Since UPC exploits the shared memory architecture (comparable to OpenMP),
we see that it is a good fit for the Intel Xeon Phi. The scaling results
for the UPC implementation are better by a factor of $1.6$ compared
to the MPI implementation. We therefore conclude that UPC is a viable
option for programming accelerators.

\subsection{Productivity}

Regarding the productivity of developing code with UPC as opposed
to MPI, there two considerations. First, how much effort is required
to obtain a working parallelization which at least scales to a couple
of nodes. Second, how difficult is it to improve such a program to
match or even exceed the performance of MPI. In the following we will
discuss these two points.

Regarding the increasing difficulty of the more performant implementations
in UPC, we identify the three versions \textit{naive}, \textit{pointer}
and \textit{barrier} as significantly more productive with respect
to code development. The development effort of these parallelizations
is comparable to OpenMP (and thus is significantly less involved than
writing an MPI program). 

If the developer is interested in strong scaling, the \textit{barrier}
version already achieves a speedup of $28.8$/$26.6$ on 64 cores
and $58.8$/$52.3$ on 256 cores on a modern system (LEO3E/LEO3).
This already exceeds the theoretical performance of OpenMP, since
this is limited by the number of cores on a node (in this case 20/12).
Compared with MPI, these speedups may be a significant margin away
from the ideal speedup of $64$ and $256$ respectively (which can
be achieved with an optimized MPI or UPC version). 

In the case of weak scaling, we can, for example, run the \textit{barrier}
version on 64 and 256 cores on the LEO3E/LEO3. For 64 cores, the runtime
increases by a factor of $1.4$/$1.4$ and for 256 cores, it increases
by a factor of $2.5$/$3.0$ compared to the serial implementation.
This means that on 64 cores we can solve, in the same amount of time,
a problem which is $2.3$/$3.8$ times as large as an ideal OpenMP
implementation would admit (which can use at most 20/12 cores). For
256 cores we can solve a problem which is $4.3$/$8.5$ times as large
as an ideal OpenMP implementation is able to handle in the same time.

Therefore, we conclude that on modern hardware we can obtain a significant
speedup compared to any implementation that only operates on a single
node, while the development effort is comparable to OpenMP (significantly
less of what would be required to do an MPI implementation). In addition,
UPC enables us to further optimize our code in order to obtain performance
that is comparable to MPI (this is done for the \textit{halo} and
\textit{patch} implementations). Of course, this requires additional
development effort. In that respect, a major advantage concerning
productivity, in our opinion, is the incremental approach for parallelism
in UPC. The possibility of debugging only small changes leads to a
significant reduction in development time for the different versions
of the program.

\section{Conclusion }

In this paper, we investigate the usability of the PGAS language UPC
for a scientific code and its competitiveness with MPI. We use a basic
fluid dynamics code to solve the Euler equations of gas dynamics and
parallelized it with both MPI and different optimization stages of
UPC for two different communication patterns. Then, we compare the
results on different hardware systems by conducting a strong and a
weak scaling analysis, respectively.

We find that in most cases, for the row implementation UPC exceeds
the MPI implementation. However, for the patched implementation, MPI
scales usually better.

We experience a major drawback at  VSC2, where barriers prove to be
extremely expensive. This issue is not nearly as significant for the
other hardware. Furthermore, on more than 1024 cores on  VSC3 the
efficiency of connection bundling degrades the performance. 

Despite these issues, especially the possibility of incremental parallelization
convinces us that UPC is a viable option for scientific computing
on these HPC systems.

\section{Acknowledgments}

We want to thank Paul Hargrove (Lawrence Berkeley National Laboratory)
for helping us to set up UPC on  VSC3. We also want to thank Martin
Thaler (ZID, University of Innsbruck) for helping us to set up UPC
on LEO3 and LEO3E.

This paper is based upon work supported by the VSC Research Center
funded by the Austrian Federal Ministry of Science, Research and Economy
(bmwfw). The computational results presented have been achieved in
part using the Vienna Scientific Cluster (VSC). This work was supported
by the Austrian Ministry of Science BMWF as part of the UniInfrastrukturprogramm
of the Focal Point Scientific Computing at the University of Innsbruck.

\bibliographystyle{plain}
\bibliography{cites}

\end{document}